\documentclass[conference]{IEEEtran}

\usepackage{amsmath}
\usepackage{amssymb}
\usepackage{physics}
\usepackage{stmaryrd}
\usepackage{tikz}
\usepackage{tikz-cd}
\usepackage{tikzit}
\usepackage{circuitikz}
\usepackage{hyperref}
\hypersetup{colorlinks=true, unicode=true, linkcolor=[rgb]{0.10,0.05,0.67}, citecolor=[rgb]{0.10,0.05,0.67}, filecolor=[rgb]{0.10,0.05,0.67}, urlcolor=[rgb]{0.10,0.05,0.67}}
\usepackage{tubes}
\usepackage{xsavebox}
\usepackage{cmll}
\usepackage[caption=false,font=footnotesize]{subfig}

\input{styles.tikzdefs}
\tikzstyle{discarding}=[fill=white, draw=black, shape=circle, style=upground]
\tikzstyle{smalldiscarding}=[fill=white, draw=black, style=upground, scale=0.75]
\tikzstyle{backdiscard}=[fill=white, draw=black, shape=circle, style=downground, scale=0.5]
\tikzstyle{smallbackdiscard}=[fill=white, draw=black, shape=circle, style=downground, scale=0.5]
\tikzstyle{state}=[fill=white, draw=black, style=triang, tikzit shape=rectangle]
\tikzstyle{kstate}=[fill=white, draw=black, style=kpoint, tikzit shape=rectangle]
\tikzstyle{kstateconj}=[fill=white, draw=black, style=kpoint conjugate, tikzit shape=rectangle]
\tikzstyle{kstateBIG}=[fill=white, draw=black, style=big kpoint, tikzit shape=rectangle]
\tikzstyle{effect}=[fill=white, draw=black, style=triangdag]
\tikzstyle{keffect}=[fill=white, draw=black, style=kpoint adjoint]
\tikzstyle{keffectconj}=[fill=white, draw=black, style=kpoint transpose]
\tikzstyle{morphdag}=[style=mapdag]
\tikzstyle{morph}=[style=hadamard]
\tikzstyle{WIDEmorph}=[style=hadamard, minimum width=14mm]
\tikzstyle{morphtrans}=[style=maptrans]
\tikzstyle{morphconj}=[style=mapconj]
\tikzstyle{CPMmorph}=[style=dmap]
\tikzstyle{CPMmorphconj}=[style=dmapconj]
\tikzstyle{CPMmorphdag}=[style=dmapdag]
\tikzstyle{CPMmorphtrans}=[style=dmaptrans]
\tikzstyle{CPMstate}=[fill=white, draw=black, style=triang, doubled]
\tikzstyle{CPMstateBIG}=[fill=white, draw=black, style={triang_lesssep}, doubled]
\tikzstyle{CPMkstate}=[fill=white, draw=black, style=kpoint, tikzit shape=rectangle, doubled]
\tikzstyle{CPMkstateconj}=[fill=white, draw=black, style=kpoint conjugate, tikzit shape=rectangle, doubled]
\tikzstyle{CPMkstateBIG}=[fill=white, draw=black, style=big kpoint, tikzit shape=rectangle, doubled]
\tikzstyle{CPMkeffect}=[fill=white, draw=black, style=kpoint adjoint, doubled]
\tikzstyle{CPMkeffectconj}=[fill=white, draw=black, style=kpoint transpose, doubled]
\tikzstyle{UHfB}=[fill=white, draw=black, style=triangdag, doubled, inner sep=-2pt]
\tikzstyle{leak}=[style=tinypoint, regular polygon rotate=-90]
\tikzstyle{leakfill}=[style=tinypoint, regular polygon rotate=-90, fill=black]
\tikzstyle{Z}=[style=dot, fill=green]
\tikzstyle{X}=[style=dot, fill=red]
\tikzstyle{black_dot}=[style=dot, fill=black]
\tikzstyle{white_dot}=[style=dot, fill=white]
\tikzstyle{link}=[fill=none, draw=none, style=dot, tikzit fill={rgb,255: red,254; green,254; blue,254}]
\tikzstyle{qblack_dot}=[style=ddot, fill=black]
\tikzstyle{qwhite_dot}=[style=ddot, fill=white]
\tikzstyle{whitephase}=[style=wphase dot, fill=white]
\tikzstyle{qredphase}=[style=phase dot, fill=red]
\tikzstyle{qgreenphase}=[style=phase dot, fill=green]
\tikzstyle{had}=[style=hadamard, doubled]
\tikzstyle{box}=[style=hadamard]
\tikzstyle{bigbox}=[style=hadamard, minimum height=4mm, minimum width=8mm]
\tikzstyle{classhad}=[style=hadamard]
\tikzstyle{antipode}=[style=anti]

\tikzstyle{dottededge}=[-, dash pattern=on 1pt off 0.7pt]
\tikzstyle{double edge}=[-, style=doubled, draw=black, tikzit draw={rgb,255: red,18; green,168; blue,191}]
\tikzstyle{arrow}=[->]
\tikzstyle{new edge style 1}=[-, draw={rgb,255: red,242; green,233; blue,206}, fill={rgb,255: red,242; green,233; blue,206}]
\tikzstyle{morphism_shade}=[-, draw=black, fill={rgb,255: red,242; green,233; blue,206}, line join=bevel]
\tikzstyle{supermap_shade}=[-, fill={rgb,255: red,216; green,215; blue,242}, draw=black, line join=bevel]
\tikzstyle{hole_shade}=[-, fill=white, draw=black, line join=bevel]
\tikzstyle{new edge style 2}=[-, draw={rgb,255: red,14; green,188; blue,83}]
\tikzstyle{new edge style 3}=[<-, draw={rgb,255: red,234; green,209; blue,255}]
\tikzstyle{new edge style 4}=[<-, draw={rgb,255: red,0; green,106; blue,106}]
\tikzstyle{new edge style 5}=[-, draw={rgb,255: red,214; green,110; blue,62}]
\tikzstyle{new edge style 6}=[-, draw={rgb,255: red,174; green,20; blue,174}]
\tikzstyle{new edge style 0}=[-, fill=none, draw={rgb,255: red,0; green,106; blue,106}]
\tikzstyle{functor box}=[-, draw={rgb,255: red,196; green,53; blue,84}]

\newcommand{\morph}[1]{\xrightarrow{#1}}

\newcommand{\pmorph}{\relbar\joinrel\mapstochar\joinrel\rightarrow}
\newcommand{\cat}[1]{\mathcal{#1}}

\newcommand{\opcat}[1]{#1^\textrm{op}}
\newcommand{\set}{\mathsf{Set}}
\newcommand{\cosmos}{\set}

\newcommand{\Prof}{\mathsf{Prof}}
\newcommand{\StProf}{\mathsf{StProf}}

\newcommand{\copsh}[1]{[#1,\cosmos]}
\newcommand{\opt}{\mathsf{Optic}}

\newcommand{\cptp}{\mathsf{CPTP}}

\newcommand{\bl}{\mathord{=}}
\newcommand{\suphom}[2]{\cat{C}(#1 \otimes-, #2 \otimes\bl)}

\newcommand{\seq}{\olessthan}
\newcommand{\Chu}{\mathsf{Chu}}
\newcommand{\Env}{\mathsf{Env}}
\newcommand{\StEnv}{\mathsf{StEnv}}
\newcommand{\true}{\mathord{\perp}}

\newcommand{\C}[1]{\cat{C}_{\vb*{#1}}}
\newcommand{\yo}[1]{y_{\vb*{#1}}}
\newcommand{\LC}[2]{LC_{\vb*{#1},\vb*{#2}}}
\newcommand{\staraut}{\ast\mathsf{Aut}}
\newcommand{\nstaraut}{\ast\mathsf{Aut}_\mathsf{N}}

\makeatletter
\newcommand{\xRightarrow}[2][]{\ext@arrow 0359\Rightarrowfill@{#1}{#2}}
\makeatother

\usepackage{cite}
\usepackage{mathtools}

\pgfsetlayers{background,strings,edgelayer,nodelayer,foreground,main}

\begin{document}

\title{A BV-Category of Spacetime Interventions}

\author{
\IEEEauthorblockN{James Hefford\IEEEauthorrefmark{1}, Matt Wilson\IEEEauthorrefmark{1}\IEEEauthorrefmark{2}}
\IEEEauthorblockA{\IEEEauthorrefmark{1}
Université Paris-Saclay, CNRS, ENS Paris-Saclay, Inria, CentraleSupélec, Laboratoire Méthodes Formelles}
\IEEEauthorblockA{\IEEEauthorrefmark{2}
University College London, PPLV Group \\ 
Email: james.hefford@inria.fr, matthew.wilson@centralesupelec.fr}
}
% \author{Anonymous Author(s)}

\newtheorem{theorem}{Theorem}
\newtheorem{lemma}{Lemma}
\newtheorem{proposition}{Proposition}
\newtheorem{definition}{Definition}
\newtheorem{remark}{Remark}
\newtheorem{example}{Example}

\maketitle

\begin{abstract}
We use the Chu construction to functorially build BV-categories from duoidal categories, demonstrating that candidate models of BV-logic can be cofreely constructed from a fragment of a model of Retoré's sequencing operator.
By using this construction to show that the strong Hyland envelope is a BV-category, we find a way to build a canonical model of spatio-temporal relationships between agents in spacetime from any symmetric monoidal category.
The concrete physical interpretation of spacetime events in this model as intervention-context pairs resolves deficiencies in previous attempts to give a general categorical semantics to quantum supermaps.
\end{abstract}

\section{Introduction}
From the perspective of abstract algebra, monoidal category theory provides a formalisation of circuit theories, that is, theories of processes with timelike and spacelike composition rules \cite{coecke_kissinger_2017}.
Along with the natural applications to quantum computer science, the existence of timelike and spacelike composition rules also forms the backbone of modern attempts to axiomatise quantum mechanics in terms of its compositional and operational features \cite{chiribella_purification, gogioso_cpt, coecke_causalcats}.
However, whilst monoidal structure captures a minimal notion of spacetime compatibility of a theory, it is not rich enough to provide a full account of the causal structure and correlations between spacetime events as conceptualised in the modern foundations of physics \cite{chiribella_switch, oreshkov, process_tensor}.

In this domain, spatio-temporal correlations are studied by modelling events in terms of the interventions that can be made by agents.
Such agents possess the ability to perform experiments by choosing the settings and viewing the outcomes of devices in their local laboratories \cite{oreshkov}.

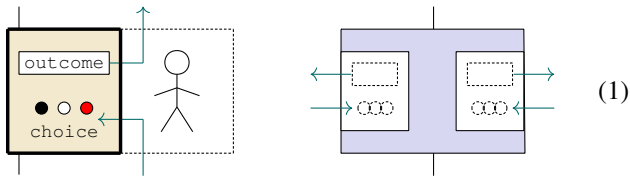
\begin{equation}\label{eq:local_lab}
	\input{figs/event_final_2.tikz} \hspace{1cm} \input{figs/context_image_2.tikz}
\end{equation}

This formulation of events in terms of the local interventions of agents motivates a dual picture of spacetime as a \textit{process-with-holes}, referred to variously as the supermap \cite{chiribella_supermaps,chiribella_circuits,chiribella_networks}, process matrix \cite{oreshkov}, or process tensor perspective \cite{process_tensor}.
These spacetime models are innately higher-order: a process-with-holes is interpreted as a context to be probed by the lower-order interventions of the local agents, thereby acting as a map on the processes of some monoidal category.

A number of important questions in the foundations of physics have prompted this conceptualisation of spacetime events.
For instance, contexts have been used to model quantum superpositions of causal structures which, if realisable, would allow for computations and correlations beyond those that can be produced in a classical spacetime \cite{chiribella_switch,chiribella_switch2}.
They have further been used to cleanly define quantum non-Markovian dynamics \cite{process_tensor}, and generalise causal modelling to the quantum setting \cite{Barrett2021}. Along with ongoing interest in the properties of correlations that can be extracted from processes-with-holes \cite{oreshkov}, there has been recent interest in both generalising these models to the study of alternative physical theories \cite{bavaresco2024indefinitecausalorderboxworld, sengupta2024achievingmaximalcausalindefiniteness, wilson_locality, sakharwade2024diagrammaticlanguagecausaloidframework, chen2024causalitydualitymultipartitegeneralized, jenvcova2024structure} and finding more abstract logical and categorical models which can adequately cope with higher-order causal structures \cite{kissinger_caus,SimmonsKissinger2022, wilson_polycategories, hefford_supermaps, bisio_2019, apadula2022nosignalling , simmons_completelogic, wilson2023mathematical , Wilson_causal, hoffreumon2022projective, jenvcova2024structure}. 

With the understanding that a fully-fledged model of spatio-temporal correlations requires us to move from processes to spacetime events, we expect that we should move away from mere monoidal categories to a more capable logical system. 
The deep inference system BV is one such logic originating in the study of extensions of multiplicative linear logic (MLL) to incorporate non-commutativity \cite{retore,retore_thesis,guglielmi,guglielmi_structures}.
BV-logic is defined in terms of the \textit{calculus of structures} and adds a non-commutative sequencing operator $\seq$ to the multiplicative conjunction $\otimes$ and disjunction $\parr$ of MLL \cite{guglielmi}.
These structures permit the following graphical depiction.

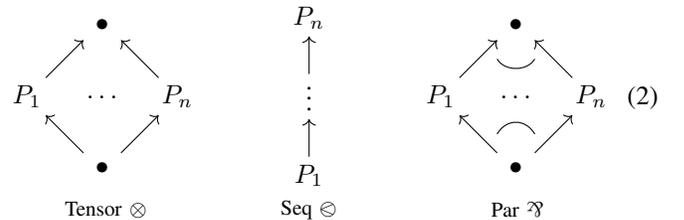
\begin{equation}\label{eq:bv_logic_structs}
	\input{figs/structures.tikz}
\end{equation}

The similarity between these structures and spacetime diagrams is particularly notable.
Indeed, the analogy between the operators $\otimes$, $\seq$ and $\parr$ and spatio-temporal correlations was appreciated in the seminal \cite{guglielmi}.
Upon interpreting the $P_i$ as the agents, we see that $\otimes$ forbids communication, $\parr$ allows arbitrary communication and $\seq$ permits just one directional communication.
Indeed, Guglielmi suggests a broad interpretation of $\seq$ as temporal composition with $P_2\seq P_1$ thought of as ``$P_1$ before $P_2$''.

\subsection{Related Work}

BV-logic has been related to causality in quantum physics in at least two ways before: by connection to the causal set approach to quantum gravity \cite{blute_quantum_causal_dynamics,blute_logical_basis}, and by connection to the study of higher-order quantum theory and indefinite causal orders \cite{kissinger_caus,SimmonsKissinger2022}.
The former \cite{blute_quantum_causal_dynamics} introduced the study of \textit{discrete quantum causal dynamics} (DQCD) to track the causal evolution of density matrices through foliations of a classical spacetime.
Spacetime events are considered to be vertices of a directed acyclic graph (DAG) with the edges representing causal links between the events.
Quantum systems are then placed over these DAGs with their evolutions compatible with the causal structure.
It was later demonstrated that the structure of these quantum causal evolutions could be interpreted as models of the system BV with the intuitive interpretation of the connectives given in equation \eqref{eq:bv_logic_structs} \cite{blute_logical_basis}.
Two concrete examples of categories which gave semantics to this logical syntax were given: the probabilistic and quantum coherence spaces of \cite{girard_coherence_spaces}.
DQCD is, however, restricted to first order: it tracks the evolution of states through spacetime but does not carry a notion of event as an intervention or a hole, and likewise does not carry a notion of higher-order process.

The other notable connection between BV-logic, causality and quantum theory was developed in \cite{kissinger_caus,SimmonsKissinger2022}.
This introduced the Caus-construction (as an instance of double gluing) and demonstrated that the higher-order processes of finite dimensional classical and quantum theory form BV-categories.
Within these categories are not only the expected first-order evolutions of systems but also the higher-order maps arising in the supermap framework.
What is surprising however is the specificity of the construction, with the Caus-construction only working in particular instances and not permitting the study of spatio-temporal correlations between systems from arbitrary physical theories.
This includes infinite dimensional quantum theory, in addition to a myriad of other exotic models stemming from Operational and Categorical Probabilistic Theories \cite{chiribella_purification,gogioso_cpt}.

Another approach to modelling quantum supermaps was presented in \cite{hefford_supermaps} developing on \cite{wilson_locality,wilson_polycategories}.
A general definition for higher-order transformations over any symmetric monoidal category was first presented in \cite{wilson_locality} and demonstrated to recover the usual definition in the case of finite dimensional quantum theory.
This definition was reinterpreted in \cite{hefford_supermaps} to show that the required algebraic laws are precisely those of a strong natural transformation between strong profunctors.
With this identification in place and by noting that strong endoprofunctors form a duoidal category \cite{garner,earnshaw,roman_thesis}, it was possible to endow the earlier models of \cite{wilson_locality,wilson_polycategories} with tensor products $\otimes$ and $\seq$ which faithfully model the causally ordered and non-ordered supermaps.
Lacking from this model was any richer logic: it was noted that the natural candidates for a negation or $\parr$ failed to be involutive and associative respectively, and thus failed to give a model of MLL or indeed BV.

The search for models of BV-logic leads us naturally to the study of its categorical semantics, which appears to have not been particularly deeply studied in the literature.
Although a notion of BV-category was suggested in \cite{blute_BV}, even the basic notions of BV-functors and BV-natural transformations are absent from the literature, and these are vital not only in the development of the theory of BV-categories themselves, but also in comparisons between physical models of spatio-temporal correlations exhibiting this logic.
We also note that, as far as we are aware there is yet to be a definitive notion of a model of BV-logic.
While this is not a topic we are going to address here, the BV-categories of \cite{blute_BV} can be placed on a firm algebraic footing and appear to fulfil at least the minimal of demands to interpret the system BV.
Sadly, no general methods for constructing BV-categories have appeared in the literature, and the only widely known non-trivial examples (where $\otimes$, $\seq$ and $\parr$ all differ) are those we have earlier discussed.
The closest we have found to a result in this direction appears in \cite{atkey_bv} where it is proven that the Chu construction \cite{chu,barr} sends duoidal pomonoids to BV-pomonoids.
To categorify this result would require the upgrading of pomonoids to monoidal categories while dealing with all the additional coherence data.

This leaves us with two thoughts: 1) that there should exist some deeper connection between monoidal categories and models of BV-logic, and 2) that such a connection would not only be a rich source of models of the system BV, but that this source could be used to develop the study of spatio-temporal correlations in less well understood domains such as infinite dimensional quantum theory or in generalised probabilistic theories.
Encouraged by the relevance of these thoughts to both physics and logic, we aim to study the question of how to construct from any monoidal category, its associated BV-category of events. 

\subsection{Contributions} 

We prove that the Chu construction sends closed normal duoidal categories to BV-categories, categorifying a result of \cite{atkey_bv} showing that the Chu construction sends duoidal pomonoids to BV ones.
Along the way, we reframe the definition of BV-categories to show that they are pseudomonoids in a 2-category of $\ast$-autonomous categories placing them on a firmer algebraic footing.
This allows us to define the 2-category $\mathsf{BV}$ of BV-categories, BV-functors and BV-transformations and handle all the coherence data in a concise manner.
We are then able to prove that the Chu construction is a right 2-adjoint to the forgetful functor $U$ from BV-categories to normal $\otimes$-closed, $\otimes$-symmetric duoidal categories, lifting the well-known 2-adjunction between closed symmetric monoidal categories and $\ast$-autonomous ones \cite{pavlovic_chu}.

\[\begin{tikzcd}
	\mathsf{CSNDuo} \ar[r, "\Chu", shift left=1ex, ""{name=top}]
	& \mathsf{BV} \ar[l, "U", shift left=1ex, ""{name=bot}] \ar[phantom, "\scriptstyle\top" , from=top, to=bot]
\end{tikzcd}\]

This allows us to construct cofree BV-categories over the fragment given by multiplicative conjunction $\otimes$, sequencing $\seq$ and a weak dualising operator $(-)^* = [-,\bot]$.
While it is already very well-known that $\Chu(\cat{C},\bot)$ is a model of linear logic, the interesting part here is that the tensor product $\seq$ can be lifted to give a sequencing operator on $\Chu(\cat{C},\bot)$ with the required self-duality and distributive structures with respect to the tensor $\otimes$ and par $\parr$.

With this tool at hand, we build upon the work of \cite{hefford_supermaps} by studying a modified version of the Hyland envelope \cite{hyland_envelope,shulman_envelope} which we dub the \textit{strong} Hyland envelope.
Explicitly, this category $\StEnv(\cat{C}) := \Chu(\StProf(\cat{C}),1)$ is given by taking the Chu construction of strong endoprofunctors with the identity profunctor $1:=\cat{C}(-,-)$ as the dualising object.
Since $\StProf(\cat{C})$ is normal duoidal, $\StEnv(\cat{C})$ is a BV-category by our general result.
This completes a series of canonical free and cofree constructions starting from any symmetric monoidal category $\cat{C}$.
\begin{equation*}
	\cat{C} \overset{D}{\rightsquigarrow} \cat{C}\times\opcat{\cat{C}} \overset{y}{\rightsquigarrow} \Prof(\cat{C}) \overset{N}{\rightsquigarrow} \StProf(\cat{C}) \overset{\Chu}{\rightsquigarrow} \StEnv(\cat{C})
\end{equation*}
Here, $D$ is the doubling (splicing) construction of \cite{earnshaw}; $y$ is the free cocompletion; $N$ is the normalisation of duoidal categories \cite{garner,earnshaw}; and $\Chu$ is the Chu construction \cite{chu,barr}.
Besides $\otimes$ which is lifted all the way along this series of adjunctions, all the other logical connectives ($\seq$, $\parr$ and $(-)^\ast$) alongside their necessary interaction, are (co)freely added.

We then justify our claim that $\StEnv(\cat{C})$ is not just a canonical BV-category, but a physically relevant one that develops and brings together aspects of those already in the literature.

The objects of $\StEnv(\cat{C})$ consist of intervention-context pairs. This unites the two dual views of spacetime events: an event is no longer just the local actions of an agent, nor the hole in a process, but a unification of the two.
The causal relationships between these events are captured by the tensors $\otimes$, $\seq$ and $\parr$ which act on the interventions in the following fashion.

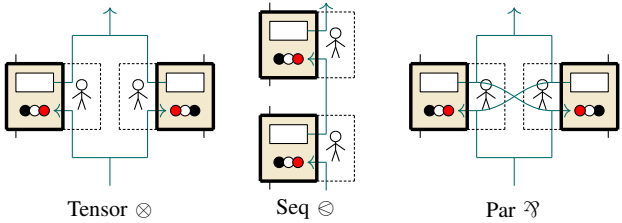
\begin{equation*}
	\input{figs/events_cartoon.tikz}
\end{equation*}

These tensors come equipped with distributors which act as,
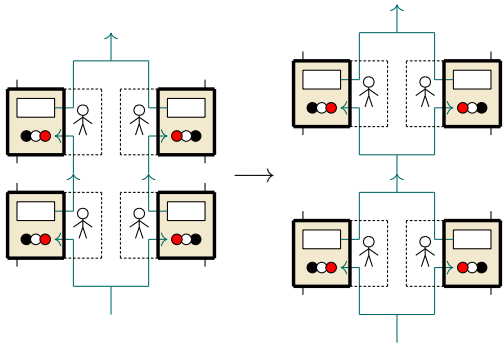
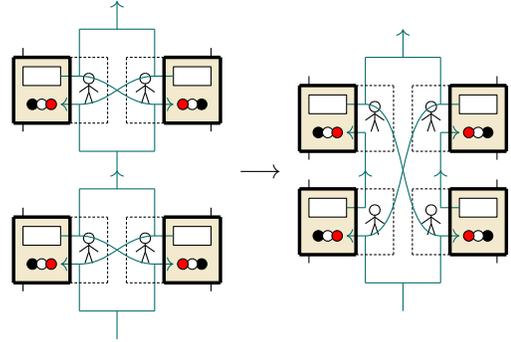
\begin{equation*} 
	\input{figs/duoidal_cartoon_1.tikz} \ \longrightarrow \ \input{figs/duoidal_cartoon_2.tikz}
	\vspace*{-3pt}
\end{equation*}

\begin{equation*}
	\input{figs/bv_cartoon_3.tikz} \ \longrightarrow \ \input{figs/bv_cartoon_4.tikz}
\end{equation*}
This improves on the weak logical structure of the model of \cite{hefford_supermaps} in a physically meaningful way: the new connective $\parr$ models the possibility of arbitrary causal communication between agents and $(-)^*$ models the duality between interventions and contexts.
We also gain the expected distributive laws between the new $\parr$ connective and the existing $\otimes$ and $\seq$.

$\StEnv(\cat{C})$ unites ideas behind both the Caus-construction and discrete quantum causal dynamics.
On the one hand, our model is manifestly higher-order like the Caus-construction, while being able to provide a BV-category of spacetime events over \textit{any} symmetric monoidal category.
Thus it can handle situations that are untreatable by the Caus-construction.
On the other hand, the similarity between the DAGs of discrete quantum causal dynamics and profunctors should be appreciated.
One can understand, at least at an intuitive level, why our model is BV by seeing the profunctors as ways of ``inflating'' the DAGs and inhabiting them with processes.

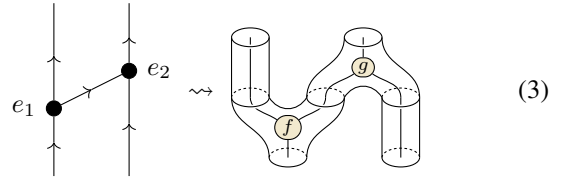
\begin{xlrbox}{inflate_dag}
	\begin{tikzpicture}[baseline=(current bounding box.center)]
		\node[copants,bot] (c) {};
		\node[pants,bot,top,anchor=leftleg] (p) at (c.rightleg) {};
		\node[tube,top,bot,anchor=bot] (t1) at (c.leftleg) {};
		\node[tube,bot,anchor=top] (t2) at (p.rightleg) {};
		\begin{pgfonlayer}{nodelayer}
			\node[label] (f) at (c.center) {$f$};
			\node[label] (g) at (p.center) {$g$};
		\end{pgfonlayer}
		\begin{pgfonlayer}{strings}
			\draw (f) to (c.belt) {};
			\draw (f) to[out=145,in=-90] (c.leftleg) {};
			\draw (f) to[out=35,in=-90] (c.rightleg) {};
			\draw (g) to (p.belt) {};
			\draw (g) to[out=-145,in=90] (p.leftleg) {};
			\draw (g) to[out=-35,in=90] (p.rightleg) {};
			\draw (t1.bot) to (t1.top) {};
			\draw (t2.bot) to (t2.top) {};
		\end{pgfonlayer}
	\end{tikzpicture}
\end{xlrbox}

\begin{equation}\label{eq:inflate}
	\input{figs/dag.tikz} \ \rightsquigarrow \ \xusebox{inflate_dag}
\end{equation}
At the same time, the higher-order transformations that are present in $\StEnv(\cat{C})$ appear to be lacking from discrete quantum causal dynamics.

The interventions and contexts that can be studied within $\StEnv(\cat{C})$ can take many exotic forms.
This includes objects like that depicted in equation \eqref{eq:inflate}; bipartite interventions modelling an agent probing an experimental device; comb-like contexts; and arbitrary black-box supermaps like the quantum switch.
One benefit of this flexibility is that one can demand decompositional properties of higher-order transformations by studying certain intervention-context pairs.
For instance, we can ensure that all single-party supermaps possess a factorisation as a comb without requiring any properties of our starting symmetric monoidal category.
This has wider effects, like demanding that multi-partite higher-order transformations possess local comb decompositions.
This idea was considered in \cite{wilson_polycategories} as a way of defining a $\parr$ in $\StProf(\cat{C})$ and here we see the strong Hyland envelope providing a cleaner and more general route to the same resolution, furthermore upgrading it to a model with logical connectives. 

\section{Linear Logic and $\ast$-autonomous Categories}
Let us start with a brief recap of linear logic and its categorical semantics.
\begin{definition}
	A symmetric monoidal category $(\cat{C},\otimes,i)$ is \textit{$\ast$-autonomous} when it is equipped with a full and faithful functor $(-)^*:\opcat{\cat{C}}\morph{}\cat{C}$ inducing a natural isomorphism $\cat{C}(a\otimes b,c) \cong \cat{C}(a,(b\otimes c)^*)$.
\end{definition}

$\ast$-autonomous categories are models of multiplicative linear logic (MLL) \cite{seely_aut} with multiplicative disjunction given by a second induced tensor product $a\parr b = (a^* \otimes b^*)^*$ with unit $i_\parr = i_\otimes^*$. Such categories are furthermore closed, with $[a,b] = a^* \parr b = (a^{**} \otimes b^*)^*$. 
A $\ast$-autonomous category is \textit{isomix} when $i_\otimes \cong i_\parr$.

\begin{definition} 
	Let $\cat{C}$ and $\cat{D}$ be $\ast$-autonomous categories.
	Let $F:\cat{C}\morph{}\cat{D}$ be a functor which is $\otimes$-lax with laxators $(l,l_0)$.
	$F$ is a \textit{$\ast$-functor} if it is additionally equipped with a natural isomorphism $s:F(-^*)\cong (F-)^*$
	%  such that $(Fa^*)^*\morph{s_{a^*}} (Fa^{**}) \cong Fa$ is equal to $(Fa^*)^*\morph{s_{a}^*}(Fa)^{**} \cong Fa$ and 
	such that the following square commutes where $\varepsilon$ is the cap.
	\begin{equation}\label{eq:pentagon1}
		\begin{tikzcd}
			(Fa)^* \otimes Fa \ar[d,"\varepsilon"'] \ar[r,"s\otimes 1"] & Fa^* \otimes Fa \ar[r,"l"] & F(a^*\otimes a) \ar[d,"F\varepsilon"] \\
			i_\parr & (Fi_\otimes)^* \ar[l,"l_0^*"] & F(i_\parr) \ar[l,"s^{-1}"]
		\end{tikzcd}
	\end{equation}
	A $\ast$-functor is \textit{normal} when its underlying $\otimes$-lax functor is normal so that it coherently preserves the unit object $F(i_\cat{C})\cong i_\cat{D}$.
\end{definition}

The reader familiar with $\ast$-autonomous categories may be surprised by the coherence condition \eqref{eq:pentagon1} which goes beyond what is usually demanded of a functor between $\ast$-autonomous categories, see e.g.\ \cite{pavlovic_chu}.
This is because the correct notion of a functor between $\ast$-autonomous categories can depend somewhat on perspective.
If we view our $\ast$-autonomous category as a linearly distributive category \cite{cockett_distributive_cats}, we might then work with the linear functors.
\begin{definition}
	A linear functor $F:\cat{C}\morph{}\cat{D}$ between symmetric linearly distributive categories is a pair of functors $F_\otimes,F_\parr:\cat{C}\morph{}\cat{D}$ such that $F_\otimes$ is $\otimes$-lax and $F_\parr$ is $\parr$-colax.
	These functors are equipped with linear strengths
	\begin{equation*} 
		F_\otimes(a\parr b) \morph{\bar{l}_\parr} F_\parr a \parr F_\otimes b, \hspace{0.75cm} F_\otimes a \otimes F_\parr b \morph{\bar{l}_\otimes} F_\parr(a\otimes b)
	\end{equation*}
	that satisfy a series of coherence conditions \cite{cockett_distributive_functors}.
	A linear functor is \textit{degenerate} (also known as Frobenius) when $F_\otimes=F_\parr$ and the strengths coincide with the laxators, $\bar{l}_\otimes=l_\otimes$ and $\bar{l}_\parr=l_\parr$.
\end{definition} 

It is known that $\otimes$-lax functors between $\ast$-autonomous categories are equivalent to linear functors \cite{cockett_distributive_functors}.
In particular, for any $\otimes$-lax $F$, $(F,(F-^*)^*)$ is always a linear functor.
Conversely, given a linear functor $(F_\otimes,F_\parr)$, the induced linear functor $(F_\otimes, (F_\otimes-^*)^*)$ is linearly equivalent to $(F_\otimes,F_\parr)$.
This induces a natural isomorphism $F_\otimes-^* \cong (F_\parr -)^*$ satisfying some coherence conditions.
We will be particularly interested in degenerate linear functors for the purposes of defining BV-categories, and so to recover these we need both the natural isomorphism $s$ \textit{and} suitable coherences captured by \eqref{eq:pentagon1}.

\begin{theorem}\label{thm:star_functor}
	A functor $F:\cat{C} \rightarrow \cat{D}$ between $\ast$-autonomous categories is degenerate linear if and only if it is a $\ast$-functor.
\end{theorem}
\begin{IEEEproof}
	In Appendix \ref{app:star_functor}.
\end{IEEEproof}

\begin{definition}
	A \textit{$\ast$-natural transformation} $\eta:F\Rightarrow G$ between $\ast$-functors $F$ and $G$, is a $\otimes$-monoidal natural transformation such that the following square commutes.
	\[\begin{tikzcd}
      {F(a)^*} & {G(a)^*} \\
      {F(a^*)} & {G(a^*)} 
      \ar["{\eta_{a^*}}"', from=2-1, to=2-2]
      \ar["{s^{F}_a}"', from=1-1, to=2-1]
      \ar["{s^{G}_{a}}", from=1-2, to=2-2]
      \ar["{\eta_{a}^*}"', from=1-2, to=1-1]
    \end{tikzcd}\]
\end{definition}

$\ast$-autonomous categories, $\ast$-functors and $\ast$-natural transformations assemble into a 2-category $\staraut$.
This 2-category is monoidal under the product of categories.
There is a monoidal sub-2-category $\nstaraut$ given by restricting to just the normal $\ast$-functors.

\subsection{The Chu Construction}\label{sec:chu}
The Chu construction provides a way of building a $\ast$-autonomous category from any closed symmetric monoidal category with sufficient pullbacks \cite{chu,barr}.
In doing so, it provides a method for turning the fragment of MLL given by multiplicative conjunction and a non-involutive $\ast$ operator, into a full model of MLL.

\begin{definition}
	Let $(\cat{C},\otimes,i)$ be a closed symmetric monoidal category (CSMC) with pullbacks and a specified object $\perp$.
	The category $\Chu(\cat{C},\true)$ has objects $(a,a',r:a\otimes a'\morph{}\true)$, consisting of a pair of objects $a$ and $a'$ of $\cat{C}$ and a chosen morphism $r:a\otimes a'\morph{}\true$.
	A morphism $(a,a',r)\morph{}(b,b',s)$ is a pair $(f,f')$ where $f:a\morph{}b$ and $f':b'\morph{}a'$ with the following diagram commuting.
	\begin{equation}\label{eq:chu_morphism}
		\begin{tikzcd}
			a\otimes b' \ar[r,"1\otimes f'"] \ar[d,"f\otimes 1"'] & a\otimes a' \ar[d,"r"] \\
			b\otimes b' \ar[r,"s"'] & \true
		\end{tikzcd}
	\end{equation}
\end{definition}

We can think of an object $(a,a',r)$ as a pair of an object $a$ and its dual $a'$ together with a chosen ``inner product'' $r$.
Such an object is known as a \textit{Chu space}.

There is a tensor product induced on $\Chu(\cat{C},\true)$ which acts on objects as,
\begin{equation}\label{eq:tensor_chu}
  (a,a',r) \otimes (b,b',s) := (a\otimes b, [a,b']\times_{[a\otimes b,\true]} [b,a'],u).
\end{equation}
The second component is the following pullback in $\cat{C}$.
\begin{equation*}
  \begin{tikzcd}
    {[a,b']\times_{[a\otimes b,\true]} [b,a']} \ar[r] \ar[d] \ar[rd,phantom,"\lrcorner"{near start}] & {[b,a']} \ar[d] \\
    {[a,b']} \ar[r] & {[a\otimes b,\true]}
  \end{tikzcd}
\end{equation*}
The vertical arrow $[b,a']\morph{}[a\otimes b,\true]$ here is given by the adjunct (under the closure of $\cat{C}$) of 
\begin{equation*}
  a\otimes b \otimes [b,a'] \morph{1\otimes\text{ev}} a\otimes a' \morph{r} \true
\end{equation*}
The horizontal arrow is similar but involving $s$.
The arrow $u$ in \eqref{eq:tensor_chu} is the adjunct of the map given by the diagonal of the pullback composed with the evaluation map,
\begin{equation*}
  a\otimes b \otimes [a,b']\times_{[a\otimes b,\true]} [b,a'] \morph{} a\otimes b\otimes [a\otimes b,\true] \morph{\text{ev}} \true
\end{equation*}
The unit of the tensor product is $(i,\true,\lambda_{\true}:i\otimes \true \morph{} \true)$.
There is a dualising functor 
\begin{equation*}
  \opcat{\Chu(\cat{C},\true)}\morph{}\Chu(\cat{C},\true):: (a,a',r) \mapsto (a',a,r\sigma),
\end{equation*}
and together with the earlier tensor one can show that $\Chu(\cat{C},\true)$ is $\ast$-autonomous \cite{chu,barr}.
There is also an evident functor
\begin{equation*}
  \cat{C} \morph{} \Chu(\cat{C},\true) :: a\mapsto (a,[a,\true],\text{ev})
\end{equation*}
which has a right adjoint given by the obvious forgetful functor.

It is known that the Chu construction is functorial $\Chu:\mathsf{CSMC}^\lrcorner_\bot \morph{} \staraut$ where $\mathsf{CSMC}^\lrcorner_\bot$ is the 2-category of closed symmetric monoidal categories with pullbacks and a specified object $\bot$ \cite{pavlovic_chu}.
The morphisms are the lax monoidal functors $F:\cat{C}\morph{}\cat{D}$ which laxly preserve the specified objects so that there exists an arrow $F(\bot_\cat{C})\morph{}\bot_\cat{D}$.
The 2-cells are the natural isomorphisms.
See \cite{pavlovic_chu} for the coherences that these data must satisfy.
\begin{remark} 
	The notion of $\ast$-functor in \cite{pavlovic_chu} does not include \eqref{eq:pentagon1}, though it is fairly straightforward to check that $\Chu(F)$ satisfies it for any functor $F$ in $\mathsf{CSMC}^\lrcorner_\bot$.
\end{remark}

Upon restricting to those CSMCs with pullbacks \textit{and} pushouts, $\Chu$ is a right 2-adjoint to the forgetful functor $U$.
\begin{equation}\label{eq:chu_adjunction}
	\begin{tikzcd}
		\mathsf{CSMC}^{\mathrlap{\ulcorner}\lrcorner}_\bot \ar[r, shift left=1ex, "\Chu", ""{name=top}]
		& \staraut^{\lrcorner} \ar[l, shift left=1ex, "U", ""{name=bot}] \ar[phantom, "\scriptstyle\top" , from=top, to=bot]
	  \end{tikzcd}
\end{equation}
Here we restrict $\staraut$ to just the $\ast$-autonomous categories with all pullbacks and only allow 2-cells that are isomorphisms.

Now we note a few useful properties of $\Chu$.
Firstly, there is a sub-2-category $\mathsf{CSMC}^{\lrcorner}_{\mathsf{N},\bot}$ given by restricting $\mathsf{CSMC}^\lrcorner_\bot$ to the normal functors: those which, up to coherent isomorphism, preserve both the unit objects and the specified objects, $F(\bot_\cat{C})\cong\bot_\cat{D}$.
In this setting, $\Chu(F)$ is itself normal so that the adjunction \eqref{eq:chu_adjunction} restricts to one involving $\staraut_\mathsf{N}$.

Additionally $\Chu$ behaves well with the monoidal structure of $\mathsf{CSMC}^\lrcorner_\bot$.
\begin{proposition}\label{prop:chu_monoidal}
	$\Chu$ is a strong monoidal 2-functor.
\end{proposition}
\begin{IEEEproof}
	In Appendix \ref{app:chu_strong}.
\end{IEEEproof}

\section{Constructing BV-Categories}
\subsection{Duoidal Categories}
Our starting point for constructing BV-categories will be a minimal model of the sequencing operator and its interaction with the tensor known as a duoidal category.

\begin{definition}
	A $\otimes$-closed, $\otimes$-symmetric \textit{duoidal} category $\cat{C}$ is a pseudomonoid in $\mathsf{CSMC}$, the monoidal 2-category of closed symmetric monoidal categories, lax monoidal functors and monoidal natural transformations.
\end{definition}

This means that a $\otimes$-closed, $\otimes$-symmetric duoidal category $\cat{C}$ is equipped with a closed symmetric monoidal structure $(\otimes,i_\otimes)$, together with an additional tensor $(\seq,i_\seq)$.
These tensors interact via natural transformations of the following form,
\begin{gather*}
	(a\seq b) \otimes (c\seq d) \morph{\delta_{abcd}} (a\otimes c)\seq (b\otimes d) \\
	i_\otimes \morph{\gamma} i_\otimes\seq i_\otimes, \qquad i_\seq \otimes i_\seq \morph{\mu} i_\seq, \qquad i_\otimes \morph{\nu} i_\seq
\end{gather*}
which must satisfy a series of coherence conditions which can be found in e.g.\ \cite{aguiar}.

A duoidal category is \textit{normal} when $i_\otimes \cong i_\seq$.
These are pseudomonoids in the monoidal 2-category $\mathsf{CSMC}_\mathsf{N}$.

\begin{definition}
	There are monoidal 2-categories $\mathsf{CSDuo}$ and $\mathsf{CSNDuo}$ of $\otimes$-symmetric, $\otimes$-closed (normal) duoidal categories, lax-lax monoidal functors and monoidal natural transformations
\end{definition}

\subsection{BV-Categories}
BV-categories were first defined in \cite{blute_BV} and intended as minimal models of BV-logic \cite{guglielmi}.
Since BV-logic is an extension of MLL to incorporate the sequencing connective of Retoré \cite{retore}, a BV-category $\cat{C}$ is an extension of a $\ast$-autonomous one.
We note that what we refer to as a BV-category is known as a BV-category \textit{with negation} in \cite{blute_BV}.

A pre-BV-category is a $\ast$-autonomous category $\cat{C}$ that is equipped with an additional monoidal structure $(\seq,i_\seq)$ such that $(\cat{C},\otimes,\seq)$ and $(\cat{C},\seq,\parr)$ are both duoidal categories.
Furthermore, there are natural isomorphisms making the $\seq$ self-dual,
\begin{equation*}
	(a\seq b)^* \cong a^*\seq b^*, \hspace{1cm} i_\seq \cong i_\seq^*.
\end{equation*}
These data are required to interact coherently, for instance the duoidal interchange for $(\seq,\parr)$ is the dual of that of $(\otimes,\seq)$, up to the self-duality of the $\seq$.
A pre-BV-category is a BV-category when $i_\seq\cong i_\otimes$.

\begin{definition}\cite{blute_BV}
	A pre-BV-category (with negation) is a $\ast$-autonomous category with an additional monoidal structure $(\seq,i_\seq)$ such that $-\seq-$ is a degenerate linear functor and the associator $\alpha_\seq$ is linear natural.
\end{definition}

Using our equivalence between $\ast$-functors and degenerate linear functors on $\ast$-autonomous categories we can reframe BV-categories as algebraic objects.
\begin{theorem}\label{thm:pseudomonoids}
	A pre-BV-category is a pseudomonoid in $\staraut$ and a BV-category is a pseudomonoid in $\nstaraut$.
\end{theorem}
\begin{IEEEproof}
	In Appendix \ref{app:pseudomonoids}.
\end{IEEEproof}
\begin{remark}
	It should be noted that our definition of pre-BV-category appears to be slightly stronger than that of \cite{blute_BV} since there they do not ask for any constraints on the unit object $i_\seq$.
	In our setting, it makes sense to demand that the functor $i_\seq:1\morph{}\cat{C}$ giving the unit in the pseudomonoid is itself a $\ast$-functor, and thus degenerate linear.
	The consequence of this is to give us the maps $\mu$ and $\nu$ of the duoidal structure, plus their coherence.
	This in turn gives a mixator $i_\otimes\morph{\nu}i_\seq\cong i_\seq^* \morph{\nu^*} i_\parr$ so that any pre-BV-category is mix.
	In the case of a BV-category $i_\otimes\cong i_\seq$ implies the category is isomix.
\end{remark}

With this abstraction of (pre-)BV-categories in place, there are obvious candidates for the higher cells between them.

\begin{definition}
	A pre-BV-functor is a pseudomonoid homomorphism in $\staraut$ and pre-BV-transformation is an intertwiner of pseudomonoid homomorphisms in $\staraut$.
	BV-functors and BV-transformations are defined analogously but in $\staraut_\mathsf{N}$.
	These structures assemble into monoidal 2-categories $\mathsf{pre}$-$\mathsf{BV}$ and $\mathsf{BV}$.
\end{definition}

Explicitly, a pre-BV-functor $F:\cat{C}\morph{}\cat{D}$ consists of a $\ast$-functor $F$ equipped with $*$-natural transformations $F(-)\seq F(-) \Rightarrow F(-\seq-)$ and $i^\cat{D}_\seq\morph{}F(i^\cat{C}_\seq)$.
The coherences are those of a lax-lax duoidal functor (see e.g.\ \cite{aguiar}) with additional compatibility with the $\ast$.
This means that not only is $F$ laxly $\otimes$-monoidal it is also laxly $\seq$-monoidal in a compatible fashion.

\subsection{Constructing BV-Categories via the Chu Construction}

\begin{theorem}\label{thm:preBV}
	Let $\cat{C}$ be a $\otimes$-closed, $\otimes$-symmetric duoidal category with all pullbacks and a chosen object $\bot$ which is a $\seq$-monoid.
	Then $\Chu(\cat{C},\bot)$ is a pre-BV-category and there is a 2-functor,
	\begin{equation*}
		\Chu: \mathsf{CSDuo}^\lrcorner_\bot \morph{} \mathsf{pre}\text{-}\mathsf{BV}.
	\end{equation*}
\end{theorem}
\begin{IEEEproof}
	In Appendix \ref{app:preBV}.
\end{IEEEproof}

One canonical choice of $\bot$ in the context of the previous theorem is the unit object for the sequencing tensor $i_\seq$.
Since this is trivially a $\seq$-monoid, $\Chu(\cat{C},i_\seq)$ is always a pre-BV-category.

\begin{theorem}\label{thm:BV}
	Let $\cat{C}$ be a $\otimes$-closed, $\otimes$-symmetric normal duoidal category with all pullbacks.
	Then $\Chu(\cat{C},i_\otimes)$ is a BV-category and there is a 2-functor,
	\begin{equation*}
		\Chu: \mathsf{CSNDuo}^\lrcorner \morph{} \mathsf{BV}.
	\end{equation*}
\end{theorem}
\begin{IEEEproof}
	In Appendix \ref{app:BV}.
\end{IEEEproof}

\begin{theorem}\label{thm:adjunct}
	The Chu construction is cofree from duoidal to BV-categories.
	More precisely, upon restricting to BV-categories with pullbacks and pushouts, $\Chu$ is a right 2-adjoint to the forgetful functor from BV-categories to $\otimes$-closed $\otimes$-symmetric normal duoidal categories.
	\begin{equation*}
		\begin{tikzcd}
			\mathsf{CSDuo}^{\mathrlap{\ulcorner}\lrcorner}_\bot \ar[r, shift left=1ex, "\Chu", ""{name=top}]
			& \mathsf{pre}\text{-}\mathsf{BV}^{\lrcorner} \ar[l, shift left=1ex, "U", ""{name=bot}] \ar[phantom, "\scriptstyle\top" , from=top, to=bot]
		\end{tikzcd}
		\hspace{1cm}
		\begin{tikzcd}
			\mathsf{CSNDuo}^{\mathrlap{\ulcorner}\lrcorner} \ar[r, shift left=1ex, "\Chu", ""{name=top}]
			& \mathsf{BV}^{\lrcorner} \ar[l, shift left=1ex, "U", ""{name=bot}] \ar[phantom, "\scriptstyle\top" , from=top, to=bot]
		\end{tikzcd}
	\end{equation*}
\end{theorem}
\begin{IEEEproof}
	The functor $\Chu:\mathsf{CSMC}^\lrcorner_\bot \morph{} \staraut$ is a right adjoint in $\mathsf{Mon2Cat}$ to the forgetful functor $U$ from $\ast$-autonomous categories to closed symmetric monoidal categories (see \eqref{eq:chu_adjunction} and \cite{pavlovic_chu}).
	From \cite{lucyshynwright2015relativesymmetricmonoidalclosed} the assignment $\text{PsMon}$ lifts to a 2-functor $\text{PsMon}: \mathsf{Mon2Cat} \rightarrow \mathsf{2Cat}$.
	Since 2-functors preserve adjunctions, it follows that $\text{PsMon}(\Chu)$ is right-adjoint to $\text{PsMon}(U)$.
\end{IEEEproof}

Let us now endeavour to unpack the content of Theorems \ref{thm:preBV} and \ref{thm:BV} to better understand the (pre-)BV structure of $\Chu(\cat{C},\bot)$.

The tensor $\seq$ of $\Chu(\cat{C},\bot)$ lifted from that of $\cat{C}$ acts on objects as,
\begin{equation*}
	(a,a',r) \seq (b,b',s) = (a\seq b, a'\seq b',m(r\seq s)\delta)
\end{equation*}
where the final component is the following arrow
\begin{equation*}
	(a\seq b)\otimes(a'\seq b') \morph{\delta} (a\otimes a') \seq (b\otimes b') \morph{r\seq s} \bot\seq \bot \morph{m} \bot
\end{equation*}
obtained from composing the duoidal distributor from $\cat{C}$ with $r$ and $s$, and the multiplication $m$ on the chosen dualising object $\bot$.
The unit object is $(i_\seq,i_\seq, i_\seq\otimes i_\seq \morph{\mu} i_\seq \morph{u} \bot)$ where $u$ is the unit of the $\seq$-monoid structure on $\bot$.

In the case that $\cat{C}$ is normal so that $\Chu(\cat{C},i_\otimes)$ is BV, the unit object of $\seq$ is $(i_\seq,i_\seq,i_\seq \seq i_\seq\cong i_\seq)$ which coincides up to isomorphism with the units of $\otimes$ and $\parr$, as expected.

We can prove explicitly that the sequencing operator is strictly self dual.

\begin{proposition}
	Under the assumptions of Theorem \ref{thm:preBV} or \ref{thm:BV}, the sequencing operator of $\Chu(\cat{C},\bot)$ is self dual so that $((a,a',r) \seq (b,b',s))^\ast = (a,a',r)^\ast \seq (b,b',s)^\ast$.
\end{proposition}
\begin{IEEEproof}
	\begin{IEEEeqnarray*}{rCl+x*}
		((a,a',r)\seq(b,b',s))^\ast & = & (a\seq b, a'\seq b',m(r\seq s)\delta)^\ast \\
		& = & (a'\seq b', a\seq b, m(r\seq s)\delta\sigma) \\
		& = & (a'\seq b', a\seq b,m(r\sigma\seq s\sigma)\delta) \\
		& = & (a',a,r\sigma) \seq (b',b,s\sigma) \\
		& = & (a,a',r)^* \seq (b,b',s)^* & \IEEEQEDhere
	\end{IEEEeqnarray*}
\end{IEEEproof}

The duoidal distributors between $\otimes$ and $\seq$ and between $\seq$ and $\parr$ are given by lifting $\delta$ from $\cat{C}$.
Writing $\mathfrak{a} = (a,a',r)$ for an object of $\Chu(\cat{C},\bot)$, there are distributors
\begin{align}
	(\mathfrak{a} \seq \mathfrak{b}) \otimes (\mathfrak{c} \seq \mathfrak{d}) \morph{\delta_{\mathfrak{abcd}}} (\mathfrak{a} \otimes \mathfrak{c}) \seq (\mathfrak{b} \otimes \mathfrak{d}) \label{eq:dist_env} \\
	(\mathfrak{a}\parr \mathfrak{b}) \seq (\mathfrak{c}\parr\mathfrak{d}) \morph{\varepsilon_{\mathfrak{abcd}}} (\mathfrak{a}\seq\mathfrak{c}) \parr (\mathfrak{b}\seq\mathfrak{d}) \label{eq:dist_env2}
\end{align}

To describe the former explicitly we calculate,
\begin{align*}
	(\mathfrak{a} \seq \mathfrak{b}) \otimes (\mathfrak{c} \seq \mathfrak{d}) = (ab,a'b') \otimes (cd,c'd') \\
	= (ab\otimes cd, [ab,c'd']\times_{[ab\otimes cd,\bot]} [cd,a'b'])
\end{align*}
where to save space we have suppressed $\seq$ by writing it as concatenation and neglected the final components involving $r$ and $s$.
On the other hand,
\begin{align*}
	& (\mathfrak{a} \otimes \mathfrak{c}) \seq (\mathfrak{b} \otimes \mathfrak{d}) \\
	& = (a\otimes c,[a,c']\times_{p} [c,a']) \seq (b\otimes d,[b,d']\times_{p} [d,b']) \\
	& = \big( (a\otimes c) (b\otimes d), ([a,c']\times_{p} [c,a']) ([b,d']\times_{p} [d,b']) \big)
\end{align*}
where $\times_p$ denotes the pullback.

Hence we see that the first component of \eqref{eq:dist_env} is given by $\delta_{abcd}$ from $\cat{C}$.
The second component is more involved and the desired arrow can be constructed with some diagram chasing.
See Appendix \ref{app:distributor} for more details.
The distributor \eqref{eq:dist_env2} is given by dualising \eqref{eq:dist_env}.

\subsection{Adding Additives}
When $\cat{C}$ is complete and cocomplete, $\Chu(\cat{C},\perp)$ is too with,
\begin{align*}
  \mathrm{lim}_i (a_i,a'_i,r_i) & = (\mathrm{lim}_i a_i, \mathrm{colim}_i a'_i,r) \\
  \mathrm{colim}_i (a_i,a'_i,r_i) & = (\mathrm{colim}_i a_i, \mathrm{lim}_i a'_i,r)
\end{align*}
where in the first case $r$ is given by uncurrying $\lim_i a_i \morph{} \lim_i [a'_i,\bot] \cong [\mathrm{colim}_i a'_i,\bot]$, and similarly for the latter.
Thus when $\cat{C}$ has products and coproducts so does $\Chu(\cat{C},\bot)$ and in the setting of Theorem \ref{thm:preBV} or \ref{thm:BV}, is moreover a (pre-)BV-category with additives.
We suggest calling these (pre-)MAV-categories as they are candidates for models of MAV-logic \cite{horne_mav}.

\section{Strong Profunctors, Interventions and Contexts}
\subsection{Interventions}
In this section we will introduce the mathematical tools that we will use to model contexts and interventions of agents in spacetime.
Let us start with the notion of a profunctor.

\begin{definition}
	Let $\cat{C}$ and $\cat{D}$ be categories. 
	A profunctor $P:\cat{C}\pmorph\cat{D}$ is a functor $\opcat{\cat{D}}\times\cat{C}\morph{}\set$.
\end{definition}

A profunctor $P$ is an assignment of a family of sets $P(d,c)$ over a pair of categories, such that they vary covariantly in $\cat{C}$ and contravariantly in $\cat{D}$.
They are also known as bimodules: the family of sets has a left action by $\cat{D}$ and a right action by $\cat{C}$, analogously to bimodules between rings.

In physical terms, a profunctor represents a basic model for a space which can be probed.
The contravariant variable represents the possibility to make a choice, while the covariant variable represents the possibility to observe some outcome.
The sets $P(d,c)$ consist of all the possible interventions by the agent.
We visualise $P$ as on the left of \eqref{eq:prof_intervention}.

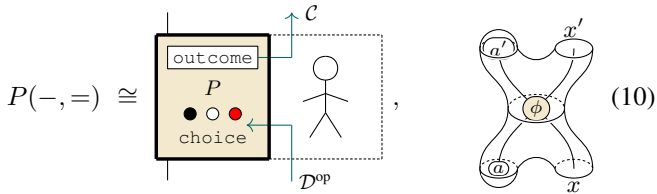
\begin{xlrbox}{intervention_basic}
	\begin{tikzpicture}[baseline=(current bounding box.center)]
		\node[copants,bot,widebelt] (c) {};
		\node[pants,bot,widebelt,anchor=belt] (p) at (c.belt) {};
		\node[bowl] at (p.leftleg) {};
		\node[cobowl] at (c.leftleg) {};
		\node[end] at (c.rightleg) {};
		\begin{pgfonlayer}{nodelayer}
			\node[label] (m) at (p.belt) {$\phi$};
			\node[representable] at (c.leftleg) {$a'$};
			\node[representable] at (p.leftleg) {$a$};
			\node[system] at ([yshift=0.7\ylength]c.rightleg) {$x'$};
			\node[system] at ([yshift=-0.6\ylength]p.rightleg) {$x$};
		\end{pgfonlayer}
		\begin{pgfonlayer}{strings}
			\draw (m) to[out=-135,in=90] (p.leftleg) {};
			\draw (m) to[out=-45,in=90] (p.rightleg) {};
			\draw (m) to[out=135,in=-90] (c.leftleg) {};
			\draw (m) to[out=45,in=-90] (c.rightleg) {};
		\end{pgfonlayer}
	\end{tikzpicture}
\end{xlrbox}

\begin{equation}\label{eq:prof_intervention}
	P(-,=) \ \cong \   \input{figs/event_prof.tikz}\ , \hspace{1cm} \xusebox{intervention_basic}
\end{equation}
An example of an intervention $P$ is drawn on the right of \eqref{eq:prof_intervention}.
Explicitly this is the profunctor $\C{a}:=\suphom{a}{a'}$, which contains for any $(x,x')$ the set of bipartite processes $\phi:a\otimes x\morph{} a'\otimes x'$.
We can think of the fixed $\vb*{a}:=(a,a')$ as the type of the device that the agent can probe and the variable $\vb*{x}:=(x,x')$ as all the possible choices $x$ and outcomes $x'$ the agent can probe $\vb*{a}$ with.
We will meet other forms of intervention later.

For the applications in this work we will only need to consider endoprofunctors $P:\cat{C}\pmorph\cat{C}$ on some given monoidal category $\cat{C}$, which together with the natural transformations between them form a category $\Prof(\cat{C})$.
This category has two tensors, the first is given by composing the profunctors in the direction of composition in $\cat{C}$,
\begin{equation*}
	(P\seq Q)(-,-) := \int^c P(-,c)\times Q(c,-),
\end{equation*}
using a special type of colimit known as a \textit{coend} \cite{loregian_coend}. 
We can visualise $\seq$ as connecting the output of $P$ into the input of $Q$, as in the left-hand diagram below.
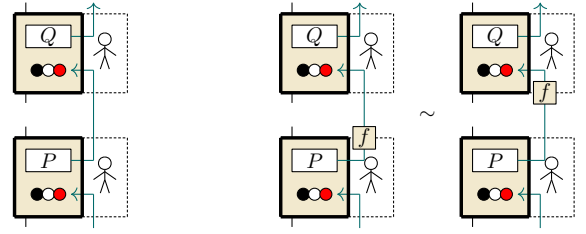
\vspace{2cm}
\begin{equation*}
	\input{figs/prof_seq_cartoon.tikz} \hspace{2cm} \input{figs/prof_seq_cartoon_quotient.tikz} 
\end{equation*}
The coend is analogous to the tensor product of bimodules: it acts to quotient the product by the action of $\cat{C}$ on the right of $P$ and the left of $Q$.
Physically, this means we can slide local maps between the two labs as in the right-hand diagram above.
The tensor product $\seq$ has the hom-profunctor $1_\cat{C}=\cat{C}(-,-)$ as its unit object.
\begin{equation*}
	1_\cat{C} \ \cong \ \input{figs/prof_iden_cartoon.tikz}
\end{equation*}
This is the space where an agent performs some trivial intervention by not interacting with the probe.

The second monoidal structure of $\Prof(\cat{C})$ is given by Day convolution \cite{day,day_thesis} over the monoidal structure of $\cat{C}$,
\begin{equation*}
	(P\otimes Q)(-,-):= \int^{aa'bb'} \substack{\cat{C}(-,a\otimes b)\times P(a,a') \\ \times Q(b,b') \times\cat{C}(a'\otimes b',-) }
\end{equation*}
which has unit object $y_i\times y^i = \cat{C}(-,i)\times\cat{C}(i,-)$.
We can think of $P\otimes Q$ as the space of separable interventions by the two agents as depicted below.
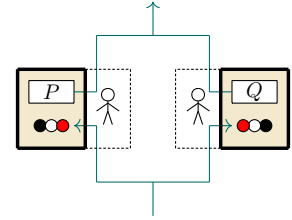
\[  \input{figs/prof_tensor_cartoon.tikz}  \]

The tensors $(\otimes,y_i\times y^i)$ and $(\seq,1_\cat{C})$ make $\Prof(\cat{C})$ a $\otimes$-closed duoidal category \cite{garner,earnshaw}.

\subsubsection{Monoidal Interventions}
The model of interventions as profunctors presented so far is lacking one key feature: compatibility between the interventions and the monoidal structure of $\cat{C}$.
This is captured by the notion of a \textit{strong} profunctor also known as a Tambara module.

\begin{definition}
	Let $\cat{C}$ be a symmetric monoidal category.
	An endoprofunctor $P:\cat{C}\pmorph\cat{C}$ is strong when it is equipped with a family of natural transformations
	\begin{equation*}
		\zeta_{abcd}: \cat{C}(a,b)\times P(c,d) \morph{} P(a\otimes c,b\otimes d)
	\end{equation*}
	such that a series of coherence conditions hold making these natural transformations essentially associative and unital \cite{tambara,pastro_street}.
\end{definition}

Strong profunctors are those which have not only actions by $\cat{C}$ in the ``vertical'' direction of composition, but also actions by $\cat{C}$ in the ``horizontal'' direction of the tensor product.
We can therefore think of them as bimodules in both dimensions simultaneously and compatibly.
From a physical perspective, the strength of $P$ means that if one agent is probing $P$ and another is performing trivial interventions, then there is an arrow mapping the agents into the same lab.

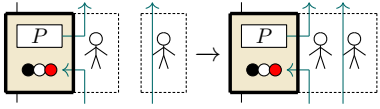
\begin{equation*}
	\input{figs/prof_strength_cartoon1.tikz} \morph{} \input{figs/prof_strength_cartoon2.tikz}
\end{equation*}
This witnesses the fact that the collection of interventions that can be performed by only one agent using device $P$ are strictly weaker than those that can be performed if both agents had access to $P$.

The natural transformations between profunctors can be straightforwardly specialised to play well with strengths.
\begin{definition}
	Let $P$ and $Q$ be strong endoprofunctors on a symmetric monoidal category $\cat{C}$.
	A natural transformation $\eta:P\Rightarrow Q$ is strong when it commutes with the strengths,
	\begin{equation*}
		\begin{tikzcd}
			\cat{C}(a,b) \times P(c,d) \ar[r,"\zeta^P_{abcd}"] \ar[d,"1\times\eta_{cd}"'] & P(a\otimes c, b\otimes d) \ar[d,"\eta_{a\otimes c,b\otimes d}"] \\
			\cat{C}(a,b) \times Q(c,d) \ar[r,"\zeta^Q_{abcd}"'] & Q(a\otimes c, b\otimes d)
		\end{tikzcd}
	\end{equation*}
\end{definition}

Strong endoprofunctors on $\cat{C}$ and strong natural transformations assemble into a category $\StProf(\cat{C})$.
This category inherits the tensors of $\Prof(\cat{C})$.
$(\seq,1_\cat{C})$ becomes a tensor by equipping $P\seq Q$ with a canonical strength induced by those of $P$ and $Q$.
The other tensor product, which we will also denote by $\otimes$, arises by a quotient of that of $\Prof(\cat{C})$ to turn it into a bimodular tensor but in the direction of the tensor product of $\cat{C}$.
See \cite{garner,earnshaw,hefford_supermaps} for further discussion of this.
The unit object for this tensor is now $1_\cat{C}$ witnessing the fact that,
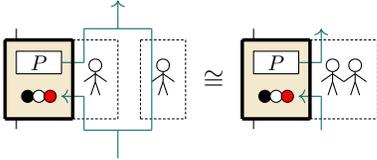
\begin{equation*}
	\input{figs/stprof_tensor_cartoon1.tikz} \ \cong \ \input{figs/stprof_tensor_cartoon2.tikz}
\end{equation*}

Altogether, $\StProf(\cat{C})$ is a $\otimes$-closed, normal duoidal category when equipped with $(\otimes,1_\cat{C})$ and $(\seq,1_\cat{C})$ \cite{garner,earnshaw}.

\subsection{Contexts}
Thus far we have seen how strong endoprofunctors provide a model of interventions by agents.
It turns out we can also use $\StProf(\cat{C})$ to model the dual picture, that of the spacetime contexts surrounding the agents.
Rather than imagining $P$ as the intervention in equation \eqref{eq:prof_intervention} we suppose $P$ takes the following form.

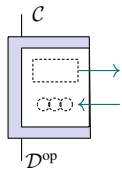
\begin{equation*}
	\input{figs/prof_context.tikz}
\end{equation*}

It is worth noticing that in this dual view, the roles of the green probing arrows and the black wires have reversed, with the former now fixed and the latter varying over the categories.

Examples of strong endoprofunctors of this form arise from the following category of \textit{coend optics} \cite{pastro_street}.

\begin{definition}
	Let $(\cat{C},\otimes,i)$ be a symmetric monoidal category.
	The category $\opt(\cat{C})$ has objects given by pairs $\vb*{a}:=(a,a')$ of objects of $\cat{C}$.
	The homs are given by the following sets,
	\begin{equation*}
	  \opt(\cat{C})(\vb*{a},\vb*{b}):= \int^x \cat{C}(b,x\otimes a)\times\cat{C}(x\otimes a',b').
	\end{equation*}
\end{definition}

The category $\opt(\cat{C})$ has morphisms given by the single-holed combs built from morphisms of $\cat{C}$.

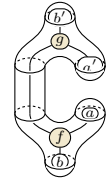
\begin{xlrbox}{optic}
	\begin{tikzpicture}[baseline=(current bounding box.center),{every node/.style}={scale=0.8}]
		\node[pants,bot] (p) {};
		\node[tube,anchor=top] (t) at (p.leftleg) {};
		\node[copants,bot,anchor=leftleg] (c) at (t.bot) {};
		% \node[end] at (c.rightleg) {};
		\node[end_dot] at (t.bot) {};
		\node[end_dot] at (p.rightleg) {};
    \node[bowl] at (p.rightleg) {};
    \node[bowl] at (c.belt) {};
    \node[cobowl] at (p.belt) {};
    \node[cobowl] at (c.rightleg) {};
		\begin{pgfonlayer}{nodelayer}
			\node[label] (g) at (p.center) {$g$};
			\node[label] (f) at (c.center) {$f$};
      \node[representable] at (p.belt) {$b'$};
      \node[representable] at (p.rightleg) {$a'$};
      \node[representable] at (c.belt) {$b$};
      \node[representable] at (c.rightleg) {$a$};
		\end{pgfonlayer}
		\begin{pgfonlayer}{strings}
			\draw (g) to[out=-155,in=90] (p.leftleg) {};
			\draw (g) to[out=-25,in=90] (p.rightleg) {};
			\draw (f) to[out=155,in=-90] (c.leftleg) {};
			\draw (f) to[out=25,in=-90] (c.rightleg) {};
			\draw (g) to (p.belt) {};
			\draw (f) to (c.belt) {};
			\draw (t.top) to (t.bot) {};
		\end{pgfonlayer}
	\end{tikzpicture}
\end{xlrbox}

\begin{equation*}
	\xusebox{optic}
\end{equation*}

Explicitly, a morphism $\vb*{a}\morph{}\vb*{b}$ is given by an equivalence class $(f,g)$ of pairs of maps $f:b\morph{}x\otimes a$ and  $g:x\otimes a'\morph{}b'$ for some $x$, quotiented by the equivalence relation generated by sliding morphisms between the two halves of the comb.

It turns out that strong endoprofunctors are the copresheaf category of coend optics, $\StProf(\cat{C}) \cong \copsh{\opt(\cat{C})}$ \cite{pastro_street}.
In particular, the Yoneda embedding $y:\opt(\cat{C})\morph{}\StProf(\cat{C})$ sends $\vb*{a}$ to $\yo{a}$ which is the set of all optics with $\vb*{a}$ as input.
We can think of this as a ``hole'' at $\vb*{a}$ in the string diagrams of $\cat{C}$.
In other words $\yo{a}$ is a strong endoprofunctor modelling contexts of type $\vb*{a}$.
These contexts have a special form, in that they all decompose in terms of maps from $\cat{C}$.
There exist strong profunctors which can model generalised contexts including quantum supermaps such as the quantum switch \cite{hefford_supermaps}.
Some of these will be discussed in more detail in the next section.

\subsection{Duality Between Interventions and Contexts}
Since $\StProf(\cat{C})$ is $\otimes$-closed, there is a weak dualising functor $(-)^*:=[-,1]$.
In \cite{hefford_supermaps} it was pointed out that this dualising functor behaves as $\yo{a}^*\cong \C{a} :=\suphom{a}{a'}$, sending the optic context at $\vb*{a}$ to the set of interventions at $\vb*{a}$.
This gives a partial duality between these simple types of contexts and inventions, but note that generally $\C{a}^*\ncong\yo{a}$.

The failure of $P^{**}\cong P$ in $\StProf(\cat{C})$ (even for the most important interventions $\C{a}$ and contexts $\yo{a}$) means that there is no hope of this category being $\ast$-autonomous or indeed BV.
This prevents it from being a fully fledged model of spacetime events over $\cat{C}$.
We will see in the next section that by uniting interventions and contexts, it is possible to rectify this problem.

\section{Spacetime Events via the Strong Hyland Envelope}
\subsection{The Strong Hyland Envelope}
Hyland \cite{hyland_envelope} and later Shulman \cite{shulman_envelope} describe a category dubbed the \textit{Hyland envelope} in the latter, which involves taking the Chu construction of the category of endoprofunctors over some given category $\cat{C}$.
They present this at the level of polycategories, but it will suffice to just consider its standard categorical analogue.

\begin{definition}[Hyland Envelope]
  \begin{equation*}
    \Env(\cat{C}) := \Chu(\Prof(\cat{C}),1_\cat{C})
  \end{equation*}
\end{definition}

Note that $\Prof(\cat{C})$ is $\otimes$-closed monoidal and being a presheaf category has all pullbacks so that $\Env(\cat{C})$ is well defined.
Also note that $1_\cat{C}$ is the unit of $\seq$ so it is trivially a $\seq$-monoid.
Therefore, by Theorem \ref{thm:preBV}, $\Env(\cat{C})$ is a pre-BV-category.

We can adjust the Hyland envelope to be better suited to our intended goals of modelling spacetime events by replacing profunctors with their strong counterparts.

\begin{definition}[Strong Hyland Envelope]
  \begin{equation*}
    \StEnv(\cat{C}) := \Chu(\StProf(\cat{C}),1_\cat{C})
  \end{equation*}
\end{definition}

Since $\StProf(\cat{C})$ is normal duoidal, $\StEnv(\cat{C})$ is a BV-category by Theorem \ref{thm:BV}.

\subsection{Intervention-Context Pairs}
A Chu space $(P,P',\eta)$ in $\StEnv(\cat{C})$ consists of a pair of strong endoprofunctors $P$ and $P'$ and a strong natural transformation $\eta:P\otimes P' \morph{} 1 $.
We can think of $P$ as the intervention and $P'$ as the context, with $\eta$ giving a specified way of evaluating the context on the intervention, so that an object looks like the following.
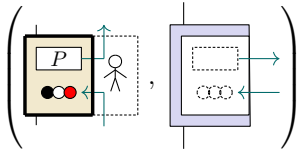
\begin{equation*}
	\left( \input{figs/intervention_tiny.tikz}\ , \ \input{figs/context_tiny.tikz} \right)
\end{equation*}

\subsection{Basic Events}

We can embed essentially all of the important structure from strong endoprofunctors into the strong envelope via the canonical functor,
\begin{equation}\label{eq:env_embedding}
	\StProf(\cat{C}) \morph{} \StEnv(\cat{C}) :: P \mapsto (P,P^*,\text{ev}).
\end{equation}
On the intervention $\C{a}$ we get the object,
\begin{xlrbox}{intervention_basic_small}
	\begin{tikzpicture}[baseline=(current bounding box.center),{every node/.style}={scale=0.8}]
		\node[copants,bot,widebelt] (c) {};
		\node[pants,bot,widebelt,anchor=belt] (p) at (c.belt) {};
		\node[bowl] at (p.leftleg) {};
		\node[cobowl] at (c.leftleg) {};
		\node[end] at (c.rightleg) {};
		\begin{pgfonlayer}{nodelayer}
			\node[label] (m) at (p.belt) {$\phi$};
			\node[representable] at (c.leftleg) {$a'$};
			\node[representable] at (p.leftleg) {$a$};
			% \node[system] at ([yshift=0.7\ylength]c.rightleg) {$x'$};
			% \node[system] at ([yshift=-0.6\ylength]p.rightleg) {$x$};
		\end{pgfonlayer}
		\begin{pgfonlayer}{strings}
			\draw (m) to[out=-135,in=90] (p.leftleg) {};
			\draw (m) to[out=-45,in=90] (p.rightleg) {};
			\draw (m) to[out=135,in=-90] (c.leftleg) {};
			\draw (m) to[out=45,in=-90] (c.rightleg) {};
		\end{pgfonlayer}
	\end{tikzpicture}
\end{xlrbox}

\begin{equation}\label{eq:events}
	(\C{a},\C{a}^*,\text{ev}) = \left( \xusebox{intervention_basic_small}\ , \ \input{figs/context_tiny2.tikz} \right)
\end{equation}
which pairs the concrete intervention $\C{a}$ with the abstract context $\C{a}^*$. 
The latter was studied in \cite{hefford_supermaps} and shown to be the space of all single-party black-box supermaps on $\vb*{a}$.
\begin{definition}
	We call the object \eqref{eq:events} an \textit{event} at $\vb*{a}$.
\end{definition}

From now on whenever we consider events we will neglect the evaluation morphisms $\text{ev}$ and leave it to the faith of the reader that those parts do indeed work out.

The functor \eqref{eq:env_embedding} is always strong $\otimes$-monoidal so that,
\begin{align*}
	(\C{a},\C{a}^*) \otimes (\C{b},\C{b}^*) \cong (\C{a}\otimes \C{b},(\C{a}\otimes\C{b})^*),
\end{align*}
essentially recovering the tensor from $\StProf(\cat{C})$.
The intervention consists of the separable maps between $\vb*{a}$ and $\vb*{b}$ \cite{hefford_supermaps}, while the context consists of all joint contexts on $\vb*{a}$ and $\vb*{b}$ that when partially evaluated in one input return a valid context on the other input.

The behaviour of the sequencing operator on events also essentially recovers that from $\StProf(\cat{C})$.
From the definition,
$(\C{a},\C{a}^*) \seq (\C{b},\C{b}^*) = (\C{a}\seq\C{b},\C{a}^*\seq\C{b}^*)$.
Here, the intervention is the sequential composition of the basic interventions at $\vb*{a}$ and $\vb*{b}$.
Similarly, the context consists of the sequential composition of all valid single-party contexts on the two parties.

Let us now consider the new connective we have accessible to us, the $\parr$.
Routine calculation shows,
\begin{equation*}
	(\C{a},\C{a}^*) \parr (\C{b},\C{b}^*) = ([\C{a}^*,\C{b}]\times_{[\C{a}^*\otimes \C{b}^*,1]} [\C{b}^*,\C{a}], \C{a}^*\otimes\C{b}^*)
\end{equation*}
The context here is the easiest to interpret: it consists of the space of single-party contexts with shared past and future.
As for the intervention, we can place bounds on it.
Firstly, for any Chu spaces $\vb*{P}=(P,P')$ and $\vb*{Q}=(Q,Q')$ the following diagram commutes,
\begin{equation*}
	\begin{tikzcd}
		\vb*{P} \parr \vb*{Q} & & \\
		& (\vb*{P} \seq \vb*{Q})\vee (\vb*{Q} \seq \vb*{P}) \ar[lu,dashed] & \vb*{P} \seq \vb*{Q} \ar[l] \ar[llu,in=0,out=145,"\tau_\parr^l"] \\
		& \vb*{Q} \seq \vb*{P} \ar[u] \ar[luu,bend left,"\tau_\parr^r"] & \vb*{P} \otimes \vb*{Q} \ar[l,"\tau_\otimes^r"] \ar[u,"\tau_\otimes^l"] \ar[lu,phantom,"\lrcorner"{near end}]
	\end{tikzcd}
\end{equation*}
where the $\tau$ are the canonical embeddings coming from the BV structure and the object in the centre is the pushout.
Taking $\vb*{P}$ and $\vb*{Q}$ to be events gives an arrow,
\begin{equation*}
	(\C{a}\seq\C{b})\vee(\C{b}\seq\C{a}) \morph{} [\C{a}^*,\C{b}]\times_{[\C{a}^*\otimes \C{b}^*,1]} [\C{b}^*,\C{a}]
\end{equation*}
from the join of the semi-localisable maps with $\vb*{a}$ and $\vb*{b}$ in either causal order into the intervention in the $\parr$ of events.

On the other hand, there is also an arrow 
\begin{equation}\label{eq:upperbound}
	[\C{a}^*,\C{b}]\times_{[\C{a}^*\otimes \C{b}^*,1]} [\C{b}^*,\C{a}] \morph{} [\C{a}^*\otimes \C{b}^*,1] \morph{} \cat{C}_{\vb*{a}\otimes\vb*{b}}
\end{equation}
where the first arrow is the diagonal of the pullback.
Noting that $\cat{C}_{\vb*{a}\otimes\vb*{b}} \cong y_{\vb*{a}\otimes\vb*{b}}^*$, the second arrow of \eqref{eq:upperbound} is the adjunct to,
\begin{equation*}
	\yo{a}\otimes\yo{b}\otimes[\yo{a}^{**}\otimes\yo{b}^{**},1] \morph{} \yo{a}^{**}\otimes\yo{b}^{**}\otimes[\yo{a}^{**}\otimes\yo{b}^{**},1] \morph{\text{ev}} 1
\end{equation*}
So we see that the intervention in the $\parr$ of events is bounded below by the join of semi-localisable maps on $\vb*{a}$ and $\vb*{b}$ and bounded above by the space of all bipartite maps on $\vb*{a}$ and $\vb*{b}$.

The functor \eqref{eq:env_embedding} is also fully faithful which allows us to recover all the higher-order transformations present in $\StProf(\cat{C})$.
For instance, a single-party higher-order map is an arrow of the form, $(\C{a},\C{a}^*) \morph{} (\C{b},\C{b}^*)$ which in turn is comprised of a pair of strong natural transformations $S:\C{a}\morph{}\C{b}$ and $S':\C{b}^*\morph{}\C{a}^*$.
$S'$ is actually determined by $S$ since diagram \eqref{eq:chu_morphism} must commute.

\subsection{First-Order Maps}
While the main focus of this article is on developing BV-logical models of spacetime interventions and higher-order processes, it is worth remarking on how the first-order maps appear in the strong Hyland envelope.

Since we interpret the object $\vb*{a}$ as the intervention at $(a,a')$ in $\cat{C}$, as a special case we can consider objects of the form and $(i,a)$ and lift these to $\StEnv(\cat{C})$.
\begin{definition}
	An event is \textit{first-order} if its carrying object is of the form $(i,a)$.
	Such events always take the form,
	\begin{equation*}
		(\cat{C}(-,a\otimes-),\cat{C}(a\otimes -,-),\text{ev}) \cong (y_{a,i},y_{i,a},\text{ev}),
	\end{equation*}
because $y_{ai}\cong \cat{C}_{ia}\cong y_{ia}^* \cong \cat{C}_{ai}^*$ and $y_{ai}^*\cong \cat{C}_{ia}^*\cong y_{ia} \cong \cat{C}_{ai}$.
\end{definition}
Note that first-order events are always causally faithful.

An arrow between first-order events,
\begin{equation*}
	(\cat{C}(-,a\otimes-),\cat{C}(a\otimes -,-)) \morph{} (\cat{C}(-,b\otimes-),\cat{C}(b\otimes -,-))
\end{equation*}
is characterised immediately by the Yoneda lemma as a morphism $a\morph{}b$ in $\cat{C}$ and is therefore a first-order transformation.

First-order events are somewhat special with regards to the tensor, the sequencing operator and the par.
\begin{lemma}
	First-order systems are closed under $\otimes, \seq$ and $\parr$.
	Furthermore, if $\vb*{P}$ and $\vb*{Q}$ are first-order systems then $\vb*{P}\parr \vb*{Q} \cong \vb*{P}\seq \vb*{Q} \cong \vb*{P}\otimes \vb*{Q}$.
\end{lemma}
\begin{IEEEproof}
	Immediate by earlier results.
\end{IEEEproof}

So far what we have outlined is the abstract sense in which the strong Hyland envelope captures key concepts in the study of causal structure and abstract spacetime events. One might wonder what the direct connection to the foundations of physics is, and there are two answers to this question. The first, is to simply note that since \eqref{eq:env_embedding} is a strong monoidal embedding of the category of strong endoprofunctors, results directly connecting strong profunctors to quantum supermaps are directly transferred. More precisely, 
arrows of the type $S:\C{a}\morph{}\C{b}$ in the category of strong endoprofunctors were considered in \cite{hefford_supermaps,wilson_locality,wilson_polycategories} and shown to recover the single-party quantum supermaps.
This extends to multi-input supermaps and the study of indefinite causal order \cite{chiribella_switch, oreshkov}: the multi-partite supermaps of \cite{hefford_supermaps} can be recovered similarly by considering maps out of $\otimes$ and $\seq$ products of objects of the form $\C{a}$ in $\StProf(\cat{C})$ and directly transferred to $\StEnv(\cat{C})$.

This first answer simply tells us that what was previously quite a well behaved model for categorical quantum supermaps can be equipped with a par to form a model of BV-logic, generalising a previously established result which builds the same conclusion from a compact closed base \cite{kissinger_caus}.
In this sense the strong Hyland envelope has advantages in for instance, being immediately applicable to infinite dimensional systems.
The second answer however, will show that the strong Hyland envelope is actually giving us much more than the category of strong endoprofunctors. New objects generated in this category have a direct interpretation in terms of \textit{local comb decompositions}, which allows for the strong Hyland envelope to cope with important deficiencies in the naive strong profunctor model of \cite{wilson_polycategories}. 

\section{Local Comb Decompositions}
One of the most natural properties one might expect of events is that given higher-order processes of type $\C{a} \rightarrow \C{a'}$ and $\C{b} \rightarrow \C{b'}$, one could construct a higher-order process of type $\cat{C}_{\vb*{a} \otimes \vb*{b}} \rightarrow \cat{C}_{\vb*{a}' \otimes \vb*{b}'}$.
In the case of finite dimensional quantum theory this has been identified as a property of the $\parr$ of finite dimensional higher-order processes \cite{kissinger_caus, bisio_2019}.

Turning to $\StProf(\cat{C})$, in the special case that $\C{a}^*\cong\yo{a}$ for all $\vb*{a}$ it is possible to partially define a $\parr$ which behaves in the desired fashion, but it cannot be extended to an associative and unital tensor on the entire category \cite{hefford_supermaps}.
The isomorphism $\C{a}^*\cong\yo{a}$ was observed to be equivalent to the claim that a category \textit{possesses decompositions for single-party supermaps}.
Indeed it holds for the category $\StProf(\cptp)$ witnessing this decomposition theorem for single-party higher-order transformations on finite dimensional quantum channels.

A more serious problem however, is that even over the category of finite dimensional unitary linear maps (which are of central importance in the study of quantum causal structure \cite{Barrett2021}), one can observe that there will be no isomorphism $\C{a}^*\cong\yo{a}$, because there exist pathological non-linear homomorphisms on strong profunctors that cannot be made to interchange over the space $\cat{C}_{\vb*{a} \otimes \vb*{b}}$ \cite{wilson_polycategories}. It turns out that there are objects in the strong Hyland envelope which encode a resolution to this problem. 

\begin{definition}
	A \textit{causally faithful} event is an object of the form $(\C{a},\yo{a},\text{ev})$.
\end{definition}
A causally faithful event, in comparison to an event, has a stronger notion of context: we demand that the only valid contexts are combs.

Since each causally faithful event $(\C{a},\yo{a})$, is dual to an object $(\yo{a},\C{a})$ in the image of the embedding \eqref{eq:env_embedding}, some of the properties of tensors of faithful events can, like for events, be seen to follow from the properties of the dualising functor and the embedding \eqref{eq:env_embedding}.

We will now unpack the concrete meaning of causally faithful events to demonstrate that they resolve the issue of the behaviour of the $\parr$.
In doing so we will also have proven that one can embed unitary supermaps into the strong Hyland envelope over the unitaries.
\subsubsection{Tensor}
Firstly, consider the tensor product of causally faithful events,
\begin{align*}
	(\C{a},\yo{a}) \otimes (\C{b},\yo{b}) = (\C{a}\otimes \C{b},[\C{a},\yo{b}] \times_{[\C{a}\otimes\C{b},1]} [\C{b},\yo{a}]).
\end{align*}
The context here is quite an interesting object as it describes precisely the set of bipartite generalised contexts at $\vb*{a}$ and $\vb*{b}$ which decompose as optics upon evaluation in one of their inputs.
We will write $\LC{a}{b}$ for this context of ``local combs''.

To see why we can think of $\LC{a}{b}$ in this way, consider a map $S$ in $[\C{a},\yo{b}](\vb{c})$.
This consists of a strong natural transformation $S:\C{a}\morph{}\yo{b}(-\otimes \vb*{c})$ which takes interventions at $\vb*{a}$ to the optic context at $\vb*{b}$.
For instance, its component $S_{\vb*{x}}$ at $\vb*{x}$ acts as,

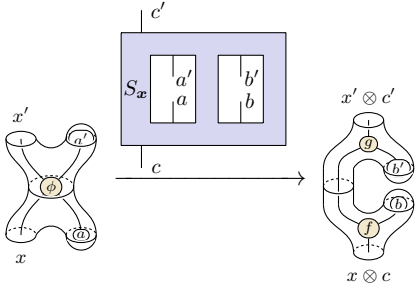
\begin{xlrbox}{supermap_basic1}
	\begin{tikzpicture}[baseline=(current bounding box.center),{every node/.style}={scale=0.8}]
		\node[copants,bot,widebelt] (c) {};
		\node[pants,bot,widebelt,anchor=belt] (p) at (c.belt) {};
		\node[bowl] at (p.rightleg) {};
		\node[cobowl] at (c.rightleg) {};
		\node[end] at (c.leftleg) {};
		\begin{pgfonlayer}{nodelayer}
			\node[label] (m) at (p.belt) {$\phi$};
			\node[representable] at (c.rightleg) {$a'$};
			\node[representable] at (p.rightleg) {$a$};
			\node[system] at ([yshift=0.8\ylength]c.leftleg) {$x'$};
			\node[system] at ([yshift=-0.8\ylength]p.leftleg) {$x$};
		\end{pgfonlayer}
		\begin{pgfonlayer}{strings}
			\draw (m) to[out=-135,in=90] (p.leftleg) {};
			\draw (m) to[out=-45,in=90] (p.rightleg) {};
			\draw (m) to[out=135,in=-90] (c.leftleg) {};
			\draw (m) to[out=45,in=-90] (c.rightleg) {};
		\end{pgfonlayer}
	\end{tikzpicture}
\end{xlrbox}

\begin{xlrbox}{optic1}
	\begin{tikzpicture}[baseline=(current bounding box.center),{every node/.style}={scale=0.8}]
		\node[pants,top] (p) {};
		\node[tube,yscale=0.75,anchor=top] (t) at (p.leftleg) {};
		\node[copants,bot,anchor=leftleg] (c) at (t.bot) {};
		% \node[end] at (c.rightleg) {};
		\node[end_dot] at (t.center) {};
		\node[end_dot] at (p.rightleg) {};
		\node[bowl] at (p.rightleg) {};
		\node[cobowl] at (c.rightleg) {};
		\begin{pgfonlayer}{nodelayer}
			\node[label] (g) at (p.center) {$g$};
			\node[label] (f) at (c.center) {$f$};
			% \node[label] (v) at (p.leftleg) {$v$};
			\node[representable] at (c.rightleg) {$b$};
			\node[representable] at (p.rightleg) {$b'$};
			\node[system] at ([yshift=0.8\ylength]p.belt) {$x'\otimes c'$};
			\node[system] at ([yshift=-0.8\ylength]c.belt) {$x\otimes c$};
		\end{pgfonlayer}
		\begin{pgfonlayer}{strings}
			\draw (g) to[out=-155,in=90] (p.leftleg) {};
			\draw (g) to[out=-25,in=90] (p.rightleg) {};
			\draw (f) to[out=155,in=-90] (c.leftleg) {};
			\draw (f) to[out=25,in=-90] (c.rightleg) {};
			\draw (g) to (p.belt) {};
			\draw (f) to (c.belt) {};
			\draw (t.top) to (t.bot) {};
		\end{pgfonlayer}
	\end{tikzpicture}
\end{xlrbox}

\begin{xlrbox}{supermap}
	\input{figs/horizontal_physics_1.tikz}
\end{xlrbox}

\begin{equation*}
	\xusebox{supermap_basic1} \ \morph{\xusebox{supermap}} \ \xusebox{optic1}
\end{equation*}

We can conceptualise $S$ as a two-input black-box supermap which returns an optic at $\vb*{b}$ when partially evaluated at $\vb*{a}$.

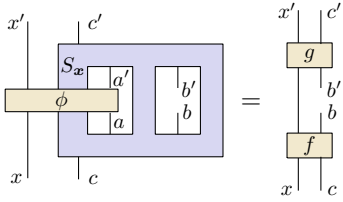
\begin{equation*}
	\input{figs/partial_eval.tikz} \ = \ \input{figs/comb.tikz}
\end{equation*}

The internal hom $[\C{a},\yo{b}]$ contains all such supermaps, and similarly for $[\C{b},\yo{a}]$ with $\vb*{a}$ and $\vb*{b}$ exchanged.
Therefore, by taking the pullback $\LC{a}{b}$ we are ensured to get just those bipartite supermaps which decompose as a comb upon evaluation in either of their inputs.

The multi-partite case follows similarly: the iterated pullback appearing in the second component of a tensor product of many causally faithful events describes precisely the space of multi-partite supermaps which decompose locally as a comb.

\subsubsection{Par} Now consider the $\parr$ of causally faithful events.
\begin{lemma}\label{lem:par}
	$(\C{a},\yo{a}) \parr (\C{b},\yo{b}) \cong (\cat{C}_{\vb*{a}\otimes\vb*{b}},y_{\vb*{a}\otimes\vb*{b}})$.
\end{lemma}
\begin{IEEEproof}
	\begin{align*}
		(\C{a},\yo{a}) \parr (\C{b},\yo{b}) & = \big( (\yo{a},\C{a}) \otimes (\yo{b},\C{b})\big)^* \\
		& = ( y_{\vb*{a}\otimes\vb*{b}}, [\yo{a},\C{b}] \times_{[\yo{a}\otimes \yo{b},1]} [\yo{b},\C{a}])^*
	\end{align*}
	where the first component follows because the $\otimes$ tensor of representable presheaves is given by the tensor of the representatives.
	Now note we have $[\yo{a},\C{b}] \cong [\yo{a}\otimes \yo{b},1] \cong [\yo{b},\C{a}]$ so that all objects involved in the pullback are naturally isomorphic, and moreover, $[\yo{a}\otimes\yo{b},1] \cong [y_{\vb*{a}\otimes\vb*{b}},1] \cong \cat{C}_{\vb*{a}\otimes\vb*{b}}$.
	Finally note that the arrows involved in the desired pullback are both isomorphisms so that the pullback is itself naturally isomorphic to $\cat{C}_{\vb*{a}\otimes\vb*{b}}$.
\end{IEEEproof}

So we see that the $\parr$ of causally faithful events acts to send the spaces $\C{a}$ and $\C{b}$ of interventions at $\vb*{a}$ and $\vb*{b}$ to the space of interventions $\cat{C}_{\vb*{a}\otimes\vb*{b}}$ on the bipartite space $\vb*{a}\otimes\vb*{b}$.
It also deals with the contexts correctly, ensuring that the spaces of optics at $\vb*{a}$ and $\vb*{b}$ are combined into the space of all optics on the joint system $\vb*{a}\otimes\vb*{b}$.
It is worth pointing out that in a category which possesses decompositions for single-party supermaps, the $\parr$ of events of course coincides with the above and thus has a simpler expression than that given earlier.

\subsubsection{Sequencing}
It is immediate to see $(\C{a},\yo{a}) \seq (\C{b},\yo{b})  \cong (\C{a}\seq\C{b},\yo{a}\seq\yo{b})$.
What is interesting here is the context, $\yo{a}\seq\yo{b}$ which is precisely the space of two-input combs.
This ensures that sequentially composed causally faithful events still possess a causally faithful context.

\subsubsection{Higher-Order Transformations}
We now turn our attention to the maps between causally faithful events.
Starting with the single-party case, suppose we have an arrow $(S,S'):(\C{a},\yo{a})\morph{}(\C{b},\yo{b})$ consisting of two strong natural transformations $S:\C{a}\morph{}\C{b}$ and $S':\yo{b}\morph{}\yo{a}$ such that the diagram \eqref{eq:chu_morphism} commutes.
$S$ is of the form of a single-party supermap from \cite{hefford_supermaps} but the compatibility of $S$ with $S'$ by \eqref{eq:chu_morphism} places extra demands of the form of this supermap.
In fact, $S'$ can be immediately characterised by the Yoneda lemma and is given by an optic $\vb*{a}\morph{}\vb*{b}$.
Since diagram \eqref{eq:chu_morphism} must commute, it is therefore necessary that $S$ be given by the action of an optic on $\C{a}$.
So we see that all single-party supermaps in $\StEnv(\cat{C})$ possess a causally faithful decomposition as an optic.

The multi-partite supermaps between causally faithful events can be characterised similarly.
For instance, an arrow $(S,S'):(\C{a},\yo{a})\otimes(\C{b},\yo{b}) \morph{} (\C{c},\yo{c})$ consists of a pair of natural transformations $S:\C{a}\otimes\C{b}\morph{}\C{c}$ and $S':\yo{c}\morph{} \LC{a}{b}$.
$S$ is a bipartite supermap from \cite{hefford_supermaps} but $S'$ is characterised by the Yoneda lemma as an element of $\LC{a}{b}(\vb*{c})$.
Since \eqref{eq:chu_morphism} must commute, we see that the higher-order transformation $(S,S')$ is a supermap possessing a local decomposition as a comb.

Similarly, $(S,S'):(\C{a}\seq\C{b},\yo{a}\seq\yo{b}) \morph{} (\C{c},\yo{c})$ consists of a pair of strong natural transformations, $S:\C{a}\seq\C{b} \morph{} \C{c}$ and $S':\yo{c}\morph{}\yo{a}\seq\yo{b}$.
The former is the same as a bipartite causally ordered supermap from \cite{hefford_supermaps} while $S'$ is characterised by the Yoneda lemma as an element of $(\yo{a}\seq\yo{b})(\vb*{c})$ which is precisely the space of two-holed optics with inputs $\vb*{a}$ and $\vb*{b}$ and output $\vb*{c}$.
Since \eqref{eq:chu_morphism} must commute, $S$ is characterised by the action of $S'$ on $\C{a}\seq\C{b}$. 

\subsubsection{Super-unitaries}
The previous discussion outlines a proof that the polycategory $\mathsf{uQS}$ of super-unitaries \cite{wilson_polycategories} embeds fully and faithfully into the model developed here. These unitary-preserving supermaps have very concrete representations in the one and two-input settings \cite{Yokojima2021consequencesof}, are used for the purpose of quantum causal modelling \cite{Barrett2021}, and are conjectured to have concrete graph-theoretic characterisations for arbitrarily many inputs \cite{vanrietvelde2023consistentcircuitsindefinitecausal}.
\begin{definition}
	The symmetric polycategory $\mathsf{uQS}$ has objects given by pairs of finite dimensional Hilbert spaces $\vb*{H} = (H,H')$ and polymorphisms $S: \vb*{H}_1 \dots \vb*{H}_n \rightarrow \vb*{K}_1 \dots \vb*{K}_m $ given by the linear maps on the tensor product of those Hilbert spaces which send unitaries to unitaries.
	This means that for tensorially extended unitaries $U_i : H_i\otimes X_i \rightarrow H_i'\otimes X_i'$, $S$ returns a unitary $S(\otimes_i U_i): \otimes_{i}(K_{i}\otimes X_i) \rightarrow \otimes_{i}(K_i' \otimes X_i')$.
	See \cite{wilson_polycategories} for more details.
\end{definition}
\begin{theorem}\label{thm:superunitaries}
	There is a fully faithful embedding of symmetric polycategories $E: \mathsf{uQS} \rightarrow \StEnv(\mathsf{U})$ from the polycategory $\mathsf{uQS}$ of super-unitaries into the strong Hyland envelope over the category $\mathsf{U}$ of finite dimensional unitaries.
\end{theorem}
\begin{IEEEproof}
In Appendix \ref{app:superunitaries}.
\end{IEEEproof}

In summary, the causally faithful objects of the strong Hyland envelope model the key concept of local comb decomposition.
The existence of such objects is quite clarifying as it demonstrates that one may have to make a choice.
On one hand, one can hold onto fully abstract spacetime contexts with events of the form $(\C{a},\C{a}^{*})$, with the sacrifice being that the $\parr$ will not allow for simultaneous application of contexts to either half of a bipartite morphism as argued for in \cite{wilson_polycategories}.
Instead the $\parr$ models an abstraction of the notion of two-way communication.
On the other hand, one may choose to study the more concrete objects $(\C{a},\yo{a})$ in which contexts have local decompositions, in this case one is rewarded by a very well behaved $\parr$ which \textit{does} support simultaneous application of contexts to bipartite morphisms.

\begin{remark}
	In \cite{wilson_polycategories} an alternative model $\mathsf{slot}$ is put forward, which holds onto partially abstract spacetime contexts by imposing bi-commutancy. The question of whether there are alternative objects within $\StEnv(\cat{C})$ with this property is left as a topic for future work. 
\end{remark}

\section{Conclusion}
The strong Hyland envelope over a symmetric monoidal category gives a canonical and well-behaved model of higher-order processes over an arbitrary physical theory. 
This work came alongside a more abstract desire to canonically construct BV-categories, and we have demonstrated that the Chu construction can cofreely achieve this while also being physically well-motivated.

While we have provided evidence that the strong Hyland envelope could be a good categorical model for the study of causal order in modern quantum foundations---in particular in the characterisation of its morphisms in familiar physical settings, and the establishment of its spatio-temporal connectives---we feel we have barely scratched the surface.

Since our model can support arbitrary symmetric monoidal categories of first-order processes, we have a candidate model for higher-order processes and indefinite causal order scenarios which includes infinite dimensional quantum systems and generalised physical theories. 
Clear next steps involve characterising this model in those new domains to produce concrete models that are more easily wieldable by those without a strong categorical background.

More concretely, natural future lines of enquiry involve understanding whether recent proposals for extending the study of indefinite causal order to specific generalised physical theories (Box-world \cite{bavaresco2024indefinitecausalorderboxworld}, and Hex-Square theory \cite{sengupta2024achievingmaximalcausalindefiniteness}) fit within the general categorical framework. 
The successful characterisation of the most basic kinds of supermaps over pure and mixed finite dimensional quantum theory should be a sign for optimism in this regard. For each generalised physical theory, the categorical framework then aims to give a stable position from which the fundamental limits on causal correlations in physical theories can be studied, likely in combination with device-independent frameworks for causal correlations and inequalities \cite{oreshkov, vanderlugt2023deviceindependentcertificationindefinitecausal, gogioso2023geometrycausality, gogioso2023topologycausality}. The framework might also be better understood if the classification of supermaps in quantum theory into subclasses can be abstracted to general monoidal categories, such subclasses include for instance quantum circuits with dynamical control of causal order \cite{wechs}, consistent circuits for indefinite causal order \cite{vanrietvelde2023consistentcircuitsindefinitecausal}, and purifiable supermaps \cite{Araujo2017purification}. 

We can also wonder what other logical features might be present in $\StEnv(\cat{C})$.
For instance, the Caus-construction has been recently demonstrated to exhibit a richer logic based on graph types which extends BV \cite{simmons_completelogic}.
It would be interesting to understand whether any of those logical features are present in the model here and whether any direct comparisons between $\mathsf{Caus}(\cat{C})$ and $\StEnv(\cat{C})$ can be made.
The Caus-construction was also shown in \cite{simmons_completelogic} to be a model of pomset logic which is known to not be equivalent to BV \cite{nguyen_pomset_bv}.
The separating statement does not appear to be generally true in our framework and we leave deeper investigation of this to future work. 

With regard to the connections between system BV and BV-categories, we would like to stress again that as far as we are aware, it has not been shown that BV-categories provide a definitive notion of a model of system BV.
We hope that our reframing of the definition of BV-categories as higher algebraic objects alongside their functors and natural transformations will be useful in this endeavour.
BV-categories can be seen as a categorification of the BV-algebras of \cite{atkey_bv} (as a special case of MAV-algebras).
There it is shown that BV is sound for BV-algebras and complete for a semantics in terms of BV-frames.
An important part of their proof is that the Chu construction can be used to send duoidal pomonoids to BV-pomonoids which Theorem \ref{thm:BV} categorifies.
It would be interesting to understand whether the remainder of the steps of their proof can be transferred over to achieve soundness and completeness results for semantics in terms of BV-categories.

Finally, we have only studied the explicit features of the Chu construction over categories of strong endoprofunctors.
This leaves open the question of the interpretation of the construction over other duoidal categories, and leads us to wonder whether the construction could be of interest in developing models for purposes other than studying spacetime interventions.

\section*{Acknowledgements}
This work has been partially funded by the French National Research Agency (ANR) within the framework of ``Plan France 2030'', under the research projects EPIQ ANR-22-PETQ-0007 and HQI-R\&D ANR-22-PNCQ-0002. The contributions of MW were split between his time at UCL and at CentraleSupélec. While based at University College London, MW was funded by the Engineering and Physical Sciences Research Council [grant number EP/W524335/1].  
The internal string ``tube'' diagrams were made using a package heavily inspired by that of \cite{bartlett_extended,bartlett_modular}.

\bibliographystyle{IEEEtran}
\bibliography{bibliography}

\begin{thebibliography}{10}
\providecommand{\url}[1]{#1}
\csname url@samestyle\endcsname
\providecommand{\newblock}{\relax}
\providecommand{\bibinfo}[2]{#2}
\providecommand{\BIBentrySTDinterwordspacing}{\spaceskip=0pt\relax}
\providecommand{\BIBentryALTinterwordstretchfactor}{4}
\providecommand{\BIBentryALTinterwordspacing}{\spaceskip=\fontdimen2\font plus
\BIBentryALTinterwordstretchfactor\fontdimen3\font minus \fontdimen4\font\relax}
\providecommand{\BIBforeignlanguage}[2]{{%
\expandafter\ifx\csname l@#1\endcsname\relax
\typeout{** WARNING: IEEEtran.bst: No hyphenation pattern has been}%
\typeout{** loaded for the language `#1'. Using the pattern for}%
\typeout{** the default language instead.}%
\else
\language=\csname l@#1\endcsname
\fi
#2}}
\providecommand{\BIBdecl}{\relax}
\BIBdecl

\bibitem{coecke_kissinger_2017}
B.~Coecke and A.~Kissinger, \emph{Picturing Quantum Processes: A First Course in Quantum Theory and Diagrammatic Reasoning}.\hskip 1em plus 0.5em minus 0.4em\relax Cambridge University Press, 2017.

\bibitem{chiribella_purification}
G.~Chiribella, G.~M. D'Ariano, and P.~Perinotti, ``Probabilistic theories with purification,'' \emph{Phys. Rev. A}, vol.~81, p. 062348, 2010.

\bibitem{gogioso_cpt}
S.~Gogioso and C.~M. Scandolo, ``Categorical probabilistic theories,'' \emph{EPTCS}, vol. 266, pp. 367--385, 2018.

\bibitem{coecke_causalcats}
B.~Coecke and R.~Lal, ``Causal categories: Relativistically interacting processes,'' \emph{Found Phys}, vol.~43, pp. 458--501, 2013.

\bibitem{chiribella_switch}
G.~Chiribella, G.~M. D'Ariano, P.~Perinotti, and B.~Valiron, ``Quantum computations without definite causal structure,'' \emph{Phys. Rev. A}, vol.~88, p. 022318, 2013.

\bibitem{oreshkov}
O.~Oreshkov, F.~Costa, and {\v C}.~Brukner, ``Quantum correlations with no causal order,'' \emph{Nature Communications}, vol.~3, no. 1092, 2012.

\bibitem{process_tensor}
\BIBentryALTinterwordspacing
F.~A. Pollock, C.~Rodr\'{\i}guez-Rosario, T.~Frauenheim, M.~Paternostro, and K.~Modi, ``Non-markovian quantum processes: Complete framework and efficient characterization,'' \emph{Phys. Rev. A}, vol.~97, p. 012127, Jan 2018. [Online]. Available: \url{https://link.aps.org/doi/10.1103/PhysRevA.97.012127}
\BIBentrySTDinterwordspacing

\bibitem{chiribella_supermaps}
G.~Chiribella, G.~M. D'Ariano, and P.~Perinotti, ``Transforming quantum operations: Quantum supermaps,'' \emph{{EPL} (Europhysics Letters)}, vol.~83, no.~3, p. 30004, 2008.

\bibitem{chiribella_circuits}
------, ``Quantum circuit architecture,'' \emph{Physical Review Letters}, vol. 101, no.~6, p. 060401, 2008.

\bibitem{chiribella_networks}
------, ``Theoretical framework for quantum networks,'' \emph{Phys. Rev. A}, vol.~80, p. 022339, 2009.

\bibitem{chiribella_switch2}
G.~Chiribella, ``Perfect discrimination of no-signalling channels via quantum superposition of causal structures,'' \emph{Phys. Rev. A}, vol.~86, p. 040301, 2012.

\bibitem{Barrett2021}
\BIBentryALTinterwordspacing
J.~Barrett, R.~Lorenz, and O.~Oreshkov, ``Cyclic quantum causal models,'' \emph{Nature Communications}, vol.~12, p. 885, 2021. [Online]. Available: \url{https://doi.org/10.1038/s41467-020-20456-x}
\BIBentrySTDinterwordspacing

\bibitem{bavaresco2024indefinitecausalorderboxworld}
\BIBentryALTinterwordspacing
J.~Bavaresco, {\"A}.~Baumeler, Y.~Guryanova, and C.~Budroni, ``Indefinite causal order in boxworld theories,'' 2024. [Online]. Available: \url{https://arxiv.org/abs/2411.00951}
\BIBentrySTDinterwordspacing

\bibitem{sengupta2024achievingmaximalcausalindefiniteness}
\BIBentryALTinterwordspacing
K.~Sengupta, ``Achieving maximal causal indefiniteness in a maximally nonlocal theory,'' 2024. [Online]. Available: \url{https://arxiv.org/abs/2411.04201}
\BIBentrySTDinterwordspacing

\bibitem{wilson_locality}
M.~Wilson, G.~Chiribella, and A.~Kissinger, ``Quantum supermaps are characterized by locality,'' 2022.

\bibitem{sakharwade2024diagrammaticlanguagecausaloidframework}
\BIBentryALTinterwordspacing
N.~Sakharwade and L.~Hardy, ``A diagrammatic language for the causaloid framework,'' 2024. [Online]. Available: \url{https://arxiv.org/abs/2407.01522}
\BIBentrySTDinterwordspacing

\bibitem{chen2024causalitydualitymultipartitegeneralized}
\BIBentryALTinterwordspacing
Y.~Chen, P.~Wang, and Z.~Wang, ``Causality and duality in multipartite generalized probabilistic theories,'' 2024. [Online]. Available: \url{https://arxiv.org/abs/2411.03903}
\BIBentrySTDinterwordspacing

\bibitem{jenvcova2024structure}
A.~Jen{\v{c}}ov{\'a}, ``On the structure of higher order quantum maps,'' \emph{arXiv preprint arXiv:2411.09256}, 2024.

\bibitem{kissinger_caus}
A.~Kissinger and S.~Uijlen, ``{A categorical semantics for causal structure},'' \emph{Logical Methods in Computer Science}, vol.~15, 2019.

\bibitem{SimmonsKissinger2022}
W.~Simmons and A.~Kissinger, ``{Higher-Order Causal Theories Are Models of BV-Logic},'' in \emph{47th International Symposium on Mathematical Foundations of Computer Science (MFCS 2022)}, ser. Leibniz International Proceedings in Informatics (LIPIcs), S.~Szeider, R.~Ganian, and A.~Silva, Eds., vol. 241.\hskip 1em plus 0.5em minus 0.4em\relax Dagstuhl, Germany: Schloss Dagstuhl -- Leibniz-Zentrum f{\"u}r Informatik, 2022, pp. 80:1--80:14.

\bibitem{wilson_polycategories}
M.~Wilson and G.~Chiribella, ``Free polycategories for unitary supermaps of arbitrary dimension,'' 2022.

\bibitem{hefford_supermaps}
J.~Hefford and M.~Wilson, ``{A Profunctorial Semantics for Quantum Supermaps},'' in \emph{Proceedings of the 39th Annual ACM/IEEE Symposium on Logic in Computer Science}, ser. LICS '24.\hskip 1em plus 0.5em minus 0.4em\relax New York, NY, USA: Association for Computing Machinery, 2024.

\bibitem{bisio_2019}
\BIBentryALTinterwordspacing
A.~Bisio and P.~Perinotti, ``Theoretical framework for higher-order quantum theory,'' \emph{Proceedings of the Royal Society A: Mathematical, Physical and Engineering Sciences}, vol. 475, no. 2225, p. 20180706, 2019. [Online]. Available: \url{http://dx.doi.org/10.1098/rspa.2018.0706}
\BIBentrySTDinterwordspacing

\bibitem{apadula2022nosignalling}
\BIBentryALTinterwordspacing
L.~Apadula, A.~Bisio, and P.~Perinotti, ``No-signalling constrains quantum computation with indefinite causal structure,'' \emph{{Quantum}}, vol.~8, p. 1241, Feb. 2024. [Online]. Available: \url{https://doi.org/10.22331/q-2024-02-05-1241}
\BIBentrySTDinterwordspacing

\bibitem{simmons_completelogic}
W.~Simmons and A.~Kissinger, ``A complete logic for causal consistency,'' 2024.

\bibitem{wilson2023mathematical}
M.~Wilson and G.~Chiribella, ``A mathematical framework for transformations of physical processes,'' 2023.

\bibitem{Wilson_causal}
\BIBentryALTinterwordspacing
------, ``Causality in higher order process theories,'' in \emph{Proceedings QPL 2021}, vol. 343.\hskip 1em plus 0.5em minus 0.4em\relax Open Publishing Association, 2021, pp. 265--300. [Online]. Available: \url{http://dx.doi.org/10.4204/EPTCS.343.12}
\BIBentrySTDinterwordspacing

\bibitem{hoffreumon2022projective}
T.~Hoffreumon and O.~Oreshkov, ``Projective characterization of higher-order quantum transformations,'' 2022.

\bibitem{retore}
C.~Retor{\'e}, ``Pomset logic: A non-commutative extension of classical linear logic,'' in \emph{Typed Lambda Calculi and Applications}, 1997, pp. 300--318.

\bibitem{retore_thesis}
------, ``{R{\'e}seaux et s{\'e}quents ordonn{\'e}s},'' Ph.D. dissertation, {Universit{\'e} Paris-Diderot - Paris VII}, 1993.

\bibitem{guglielmi}
A.~Guglielmi, ``A system of interaction and structure,'' \emph{ACM Transactions on Computational Logic}, vol.~8, no.~1, 2007.

\bibitem{guglielmi_structures}
A.~Guglielmi and L.~Stra{\ss}burger, ``Non-commutativity and mell in the calculus of structures,'' in \emph{Computer Science Logic}, L.~Fribourg, Ed., 2001, pp. 54--68.

\bibitem{blute_quantum_causal_dynamics}
R.~Blute, I.~Ivanov, and P.~Panangaden, ``Discrete quantum causal dynamics,'' \emph{International Journal of Theoretical Physics}, vol.~42, pp. 2025--2041, 2003.

\bibitem{blute_logical_basis}
R.~F. Blute, A.~Guglielmi, I.~T. Ivanov, P.~Panangaden, and L.~Stra{\ss}burger, \emph{A Logical Basis for Quantum Evolution and Entanglement}, 2014, pp. 90--107.

\bibitem{girard_coherence_spaces}
J.-Y. Girard, ``Between logic and quantic: a tract,'' \emph{Linear logic in computer science}, vol. 316, p. 346, 2004.

\bibitem{garner}
R.~Garner and I.~L. Franco, ``Commutativity,'' \emph{Journal of Pure and Applied Algebra}, vol. 220, no.~5, pp. 1707--1751, 2016.

\bibitem{earnshaw}
M.~Earnshaw, J.~Hefford, and M.~Rom{\'a}n, ``The produoidal algebra of process decomposition,'' 2023.

\bibitem{roman_thesis}
M.~Rom{\'a}n, ``Monoidal context theory,'' Ph.D. dissertation, Tallinn University of Technology, 2023.

\bibitem{blute_BV}
R.~Blute, P.~Panangaden, and S.~Slavnov, ``Deep inference and probabilistic coherence spaces,'' \emph{Applied Categorical Structures}, vol.~20, pp. 209--228, 2012.

\bibitem{atkey_bv}
R.~Atkey and W.~Kokke, ``A semantic proof of generalised cut elimination for deep inference,'' \emph{Electronic Notes in Theoretical Informatics and Computer Science}, vol. Volume 4 - Proceedings of MFPS XL, 2024.

\bibitem{chu}
P.-H. Chu, ``Constructing $\ast$-autonomous categories,'' appendix to \cite{barr}.

\bibitem{barr}
M.~Barr, ``$\ast$-autonomous categories,'' \emph{Lecture Notes in Mathematics}, vol. 752, 1979.

\bibitem{pavlovic_chu}
D.~Pavlovi{\'c}, ``Chu i: cofree equivalences, dualities and *-autonomous categories,'' \emph{Mathematical Structures in Computer Science}, vol.~7, no.~1, pp. 49--73, 1997.

\bibitem{hyland_envelope}
J.~Hyland, ``Proof theory in the abstract,'' \emph{Annals of Pure and Applied Logic}, vol. 114, no.~1, pp. 43--78, 2002.

\bibitem{shulman_envelope}
M.~Shulman, ``$\ast$-autonomous envelopes and conservativity,'' \emph{Electronic Proceedings in Theoretical Computer Science}, vol. 353, pp. 175--194, 2021.

\bibitem{seely_aut}
R.~Seely, ``Linear logic, $\ast$-autonomous categories and cofree coalgebras,'' \emph{Contemporary Mathematics}, vol.~92, 1989.

\bibitem{cockett_distributive_cats}
J.~Cockett and R.~Seely, ``Weakly distributive categories,'' \emph{Journal of Pure and Applied Algebra}, vol. 114, no.~2, pp. 133--173, 1997.

\bibitem{cockett_distributive_functors}
------, ``Linearly distributive functors,'' \emph{Journal of Pure and Applied Algebra}, vol. 143, no.~1, pp. 155--203, 1999.

\bibitem{aguiar}
S.~M. Marcelo~Aguiar, \emph{Monoidal Functors, Species and Hopf Algebras}.\hskip 1em plus 0.5em minus 0.4em\relax CRM Monograph Series, 2010, vol.~29.

\bibitem{lucyshynwright2015relativesymmetricmonoidalclosed}
\BIBentryALTinterwordspacing
R.~B.~B. Lucyshyn-Wright, ``Relative symmetric monoidal closed categories i: Autoenrichment and change of base,'' 2015. [Online]. Available: \url{https://arxiv.org/abs/1507.02220}
\BIBentrySTDinterwordspacing

\bibitem{horne_mav}
R.~Horne, ``The consistency and complexity of multiplicative additive system virtual,'' \emph{Scientific Annals of Computer Science}, vol.~25, no.~2, 2015.

\bibitem{loregian_coend}
F.~Loregian, \emph{(Co)end Calculus}, ser. London Mathematical Society Lecture Note Series.\hskip 1em plus 0.5em minus 0.4em\relax Cambridge University Press, 2021.

\bibitem{day}
B.~Day, ``On closed categories of functors,'' in \emph{Reports of the Midwest Category Seminar IV}, vol. 137.\hskip 1em plus 0.5em minus 0.4em\relax Berlin, Heidelberg: Springer Berlin Heidelberg, 1970, pp. 1--38.

\bibitem{day_thesis}
------, ``Construction of biclosed categories,'' Ph.D. dissertation, University of New South Wales, 1970.

\bibitem{tambara}
D.~Tambara, ``{Distributors on a tensor category},'' \emph{Hokkaido Mathematical Journal}, vol.~35, no.~2, pp. 379 -- 425, 2006.

\bibitem{pastro_street}
C.~Pastro and R.~Street, ``Doubles for monoidal categories,'' \emph{Theory and Applications of Categories}, vol.~21, no.~4, pp. 61--75, 2008.

\bibitem{Yokojima2021consequencesof}
\BIBentryALTinterwordspacing
W.~Yokojima, M.~T. Quintino, A.~Soeda, and M.~Murao, ``Consequences of preserving reversibility in quantum superchannels,'' \emph{{Quantum}}, vol.~5, p. 441, Apr. 2021. [Online]. Available: \url{https://doi.org/10.22331/q-2021-04-26-441}
\BIBentrySTDinterwordspacing

\bibitem{vanrietvelde2023consistentcircuitsindefinitecausal}
\BIBentryALTinterwordspacing
A.~Vanrietvelde, N.~Ormrod, H.~Kristj{\'a}nsson, and J.~Barrett, ``Consistent circuits for indefinite causal order,'' 2023. [Online]. Available: \url{https://arxiv.org/abs/2206.10042}
\BIBentrySTDinterwordspacing

\bibitem{vanderlugt2023deviceindependentcertificationindefinitecausal}
\BIBentryALTinterwordspacing
T.~van~der Lugt, J.~Barrett, and G.~Chiribella, ``Device-independent certification of indefinite causal order in the quantum switch,'' 2023. [Online]. Available: \url{https://arxiv.org/abs/2208.00719}
\BIBentrySTDinterwordspacing

\bibitem{gogioso2023geometrycausality}
\BIBentryALTinterwordspacing
S.~Gogioso and N.~Pinzani, ``The geometry of causality,'' 2023. [Online]. Available: \url{https://arxiv.org/abs/2303.09017}
\BIBentrySTDinterwordspacing

\bibitem{gogioso2023topologycausality}
\BIBentryALTinterwordspacing
------, ``The topology of causality,'' 2023. [Online]. Available: \url{https://arxiv.org/abs/2303.07148}
\BIBentrySTDinterwordspacing

\bibitem{wechs}
J.~Wechs, H.~Dourdent, A.~A. Abbott, and C.~Branciard, ``Quantum circuits with classical versus quantum control of causal order,'' \emph{PRX Quantum}, vol.~2, p. 030335, 2021.

\bibitem{Araujo2017purification}
\BIBentryALTinterwordspacing
M.~Ara{\'{u}}jo, A.~Feix, M.~Navascu{\'{e}}s, and {\v{C}}.~Brukner, ``A purification postulate for quantum mechanics with indefinite causal order,'' \emph{{Quantum}}, vol.~1, p.~10, Apr. 2017. [Online]. Available: \url{https://doi.org/10.22331/q-2017-04-26-10}
\BIBentrySTDinterwordspacing

\bibitem{nguyen_pomset_bv}
L.~T.~D. Nguy{{\^e}}n and L.~Stra{\ss}burger, ``{BV and Pomset Logic Are Not the Same},'' in \emph{30th EACSL Annual Conference on Computer Science Logic (CSL 2022)}, vol. 216, 2022, pp. 32:1--32:17.

\bibitem{bartlett_extended}
B.~Bartlett, C.~L. Douglas, C.~J. Schommer-Pries, and J.~Vicary, ``Extended 3-dimensional bordism as the theory of modular objects,'' 2014.

\bibitem{bartlett_modular}
------, ``Modular categories as representations of the 3-dimensional bordism 2-category,'' 2015.

\bibitem{day_monoidal_bicats}
B.~Day and R.~Street, ``Monoidal bicategories and hopf algebroids,'' \emph{Advances in Mathematics}, vol. 129, no.~1, pp. 99--157, 1997.

\bibitem{McCrudden2000Balanced}
P.~McCrudden, ``Balanced coalgebroids,'' \emph{Theory and Applications of Categories}, vol.~7, no.~6, pp. 71--147, 2000.

\bibitem{hefford_coend}
J.~Hefford and C.~Comfort, ``{Coend Optics for Quantum Combs},'' in \emph{Proceedings ACT 2022}, vol. 380, 2023, pp. 63--76.

\end{thebibliography}

\appendices

\section{Proofs}

\subsection{Distributor in $\Chu(\cat{C},\bot)$}\label{app:distributor}
The following diagram commutes.
\begin{equation*}
	\begin{tikzcd}[column sep=small]
\substack{ ([a,c']\times_p [b,d']) \seq \\ ([c,a']\times_p [d,b']) } \arrow[r, "\pi_1\seq\pi_1"] \arrow[dd, "\pi_2\seq\pi_2"']  & {[a,c']\seq [b,d']} \arrow[dd] \arrow[r,"\text{ten}_\seq"]     & {[ab,c'd']} \arrow[dddd] \\
 &  &        \\
{[c,a']\seq [d,b']} \arrow[r] \arrow[dd,"\text{ten}_\seq"']                                                                           & \substack{[a\otimes c,\bot]\seq \\ [b\otimes d,\bot]} \arrow[d,"\text{ten}_\seq"] &               &           \\
   &  {[(a\otimes c)(b\otimes d),\bot\bot]} \arrow[rd,"{[\delta,m]}"] &          \\
{[cd,a'b']} \arrow[rr]     &             & {[ab\otimes cd,\bot]}      
\end{tikzcd}
\end{equation*}
Here $\text{ten}_\seq$ is the canonical map that comes from currying the duoidal distributor and the evaluation morphisms,
\begin{equation*}
	([a,b]\seq[c,d])\otimes(a\seq c) \morph{\delta} ([a,b]\otimes a)\seq ([c,d]\otimes c) \morph{\text{ev}\seq\text{ev}} b\seq d
\end{equation*}
The unlabelled arrows are the canonical ones arising in the pullbacks we are working with.
The desired arrow
\begin{align*}
	[a,c']\times_{[a\otimes c,\bot]} [c,a'])\seq ([b,d']\times_{[b\otimes d,\bot]} [d,b'] \\ 
	\morph{} [ab,c'd']\times_{[ab\otimes cd,\bot]} [cd,a'b']
\end{align*}
arises from the universal property of the pullback.

\subsection{Proof of Theorem \ref{thm:star_functor}}\label{app:star_functor}
\begin{IEEEproof}
	Suppose $F$ is a $\ast$-functor.
	Since $F$ is $\otimes$-lax, $(F,F(-^*)^* := \bar{F}(-))$ is a linear functor by the results of \cite{cockett_distributive_functors}.
	The natural isomorphism $s:(Fa)^*\morph{}Fa^*$ induces a natural isomorphism $k$ between $F$ and its de Morgan dual $\bar{F}$.
	\begin{equation*}
		\input{figs/demorgan_iso.tikz}
	\end{equation*}
	So $F$ becomes $\parr$-colax with colaxator $l_\parr$ inherited from that of $\bar{F}$.
	$(F,F)$ then becomes a degenerate linear functor so long as we can lift the linear coherences of $(F,\bar{F})$ along the isomorphism $k$.

	In particular we want the outside of following diagram to commute, which it does if the three squares containing exclamation marks commute.
	\begin{equation*}
		\begin{tikzcd}[column sep=small]
			Fa \otimes F(b \parr c) \arrow[r, "1 \otimes \bar{l}_\parr"] \arrow[d, "l_\otimes"'] \ar[rr,bend left, "1\otimes l_\parr"] \ar[rr,bend left=15,"!",phantom] & Fa \otimes (\bar{F}b \parr Fc) \arrow[d, "\delta"] \ar[r,"\cong"] & Fa \otimes (Fb \parr Fc) \ar[d,"\delta"]\\
			F(a \otimes (b \parr c)) \arrow[d, "F(\delta)"'] & (Fa \otimes \bar{F}b) \parr Fc \arrow[d, "\bar{l}_\otimes \parr 1"] \ar[r,"\cong"] \ar[rd,"!",phantom] & (Fa \otimes Fb) \parr Fc \ar[d,"l_\otimes \parr 1"]\\
			F((a \otimes b) \parr c) \arrow[r, "\bar{l}_\parr"] \ar[rr,bend right,"l_\parr"] \ar[rr,bend right=15,"!",phantom] & \bar{F}(a \otimes b) \parr Fc \ar[r,"\cong"] & F(a \otimes b) \parr Fc
		\end{tikzcd}
	\end{equation*}
	So we require the following square to commute relating the laxator $l_\otimes$ and the linear strength $\bar{l}_\otimes$,
	\begin{equation}\label{eq:star_functor_proof}
		\begin{tikzcd}
			F(a)\otimes F(b) \ar[r,"1\otimes k_b"] \ar[d,"l_\otimes"'] & Fa\otimes (Fb^*)^* \ar[d,"\bar{l}_\otimes"] \\
			F(a\otimes b) \ar[r,"k_{a\otimes b}"] & (F(a\otimes b)^*)^*
		\end{tikzcd}
	\end{equation}
	and the following triangle relating the colaxator $l_\parr$ to the strength $\bar{l}_\parr$.
	\begin{equation}\label{eq:star_functor_proof2}
		\begin{tikzcd}
			F(a\parr b) \ar[r,"\bar{l}_\parr"] \ar[d,"l_\parr"'] & \bar{F}a\parr Fb \\
			Fa \parr Fb \ar[ur,"k_a\parr 1"'] &
		\end{tikzcd}
	\end{equation}

	Let us start by proving that \eqref{eq:star_functor_proof} commutes.
	The coherence \eqref{eq:pentagon1} for the $\ast$-functor $F$ has the following presentation in the string diagrams (proof nets),
	\begin{equation*}
		\input{figs/cap_eqn.tikz}
	\end{equation*}
	where the black dot is given by
	\begin{equation*}
		\input{figs/black_dot.tikz}
	\end{equation*}

	Square \eqref{eq:star_functor_proof} takes the following form where the diagram for the linear strength $\bar{l}_\otimes$ can be found in \cite{cockett_distributive_functors}.

	\begin{equation*}
		\input{figs/strength.tikz}
	\end{equation*}

	Which can be seen to hold by the following graphical manipulations.
	\begin{equation*}
		\input{figs/strength_proof1.tikz}
	\end{equation*}
	\begin{equation*}
		\input{figs/strength_proof2.tikz}
	\end{equation*}
	\begin{equation*}
		\input{figs/strength_proof3.tikz}
	\end{equation*}
	\begin{equation*}
		\input{figs/strength_proof4.tikz}
	\end{equation*}

	The triangle \eqref{eq:star_functor_proof2} takes the following form (again, see \cite{cockett_distributive_functors} for the diagrams for the linear costrength and the colaxator).

	\begin{equation*}
		\input{figs/costrength.tikz}
	\end{equation*}

	That this holds follows very similarly to the case of the laxator and linear strength.

	\begin{equation*}
		\input{figs/costrength_proof1.tikz}
	\end{equation*}
	\begin{equation*}
		\input{figs/costrength_proof2.tikz}
	\end{equation*}
	\begin{equation*}
		\input{figs/costrength_proof3.tikz}
	\end{equation*}
	\begin{equation*}
		\input{figs/costrength_proof4.tikz}
	\end{equation*}

	So we have shown that the coherence \eqref{eq:pentagon1} is sufficient to ensure that the laxator coincides with the linear strength and the colaxator with the linear costrength (up to the induced isomorphism $k$).
	As a result $(F,F)$ becomes a linear functor.

	Now we show the converse.
	So suppose that $(F,F)$ is a linear functor.
	Then $(F,\bar{F})$ is also linear and linearly equivalent to $(F,F)$, so that \eqref{eq:star_functor_proof} and \eqref{eq:star_functor_proof2} commute.
	Upon taking $b$ to be $i_\otimes$ in \eqref{eq:star_functor_proof} and composing by the laxator $l_0$ and the map $F\rho^*$ one recovers the coherence \eqref{eq:pentagon1}.
	The left hand side of \eqref{eq:star_functor_proof} becomes the following.
	\begin{equation*}
		\input{figs/cap_proof1.tikz}
	\end{equation*}
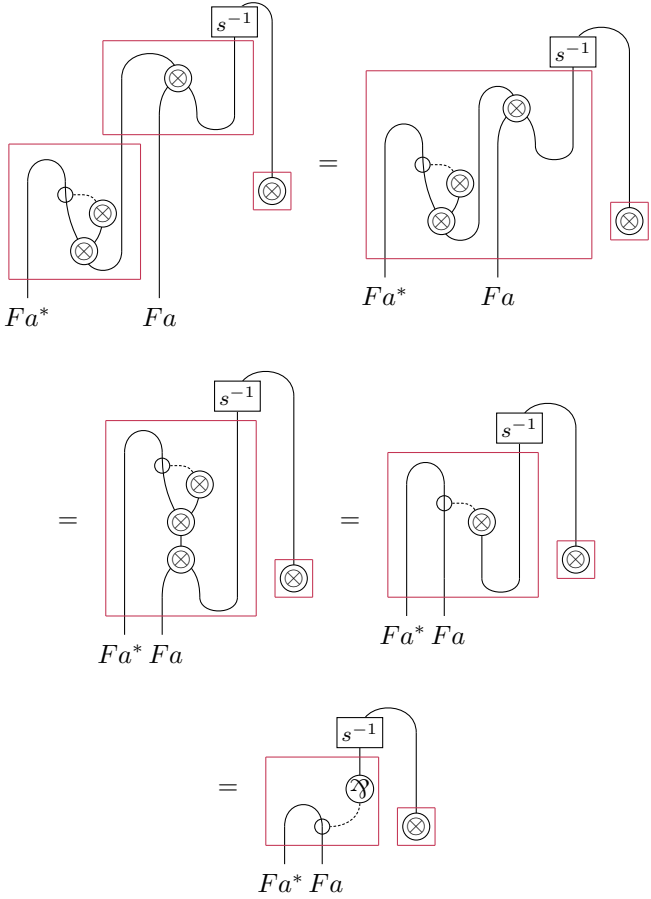
	\begin{equation*}
		\input{figs/cap_proof2.tikz}
	\end{equation*}
	\begin{equation*}
		\input{figs/cap_proof3.tikz}
	\end{equation*}

	While the right hand side of \eqref{eq:star_functor_proof} becomes the following.
	\begin{equation*}
		\input{figs/cap_proof1a.tikz}
	\end{equation*}
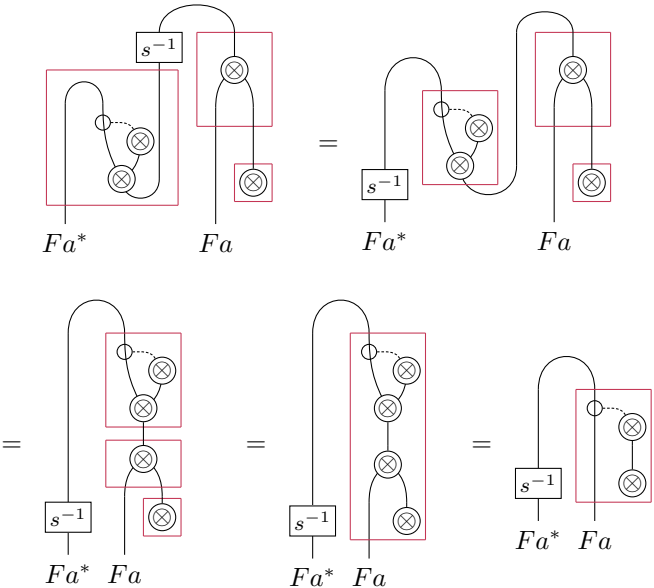
	\begin{equation*}
		\input{figs/cap_proof2a.tikz}
	\end{equation*}
	\begin{equation*}
		\input{figs/cap_proof3a.tikz}
	\end{equation*}
\end{IEEEproof}

Let us now comment on a few other notable coherence issues.
Firstly, one might be concerned that there are two canonical natural isomorphisms $Fa\cong \bar{F}a$, one given by $\bar{F}a \morph{s_a^*} (Fa)^{**} \morph{d_{Fa}} Fa$ and the other given by $\bar{F}a \morph{s_{a^*}} Fa^{**} \morph{Fd} Fa$ where $d:a^{**}\morph{}a$ is the double dualisation natural isomorphism.
In fact, these two natural isomorphisms agree for any $\ast$-functor.

\begin{lemma}
	Let $F$ be a $\ast$-functor, then the following diagram commutes.
	\begin{equation*}
		\begin{tikzcd}
			\bar{F}a \ar[r,"s_a^*"] \ar[d,"s_{a^*}"'] & (Fa)^{**} \ar[d,"d_{Fa}"] \\
			Fa^{**} \ar[r,"Fd"'] & Fa
		\end{tikzcd}
	\end{equation*}
\end{lemma}
\begin{IEEEproof}
	\begin{equation*}
		\input{figs/double_dual_proof1.tikz}
	\end{equation*}
	\begin{equation*}
		\input{figs/double_dual_proof2.tikz}
	\end{equation*}
	Bending up the right-hand input completes the proof. 
	\begin{equation*}
		\input{figs/double_dual_proof3.tikz}
	\end{equation*}
\end{IEEEproof}

\subsection{Proof of Proposition \ref{prop:chu_monoidal}}\label{app:chu_strong}

\begin{IEEEproof}
	There is an isomorphism $\Chu(\cat{C},\bot_\cat{C})\times\Chu(\cat{D},\bot_\cat{D}) \cong \Chu(\cat{C}\times\cat{D},(\bot_\cat{C},\bot_\cat{D}))$.
	The $\ast$-functor witnessing this acts on objects in the intuitive fashion,
	\begin{equation*}
		\big( (c,c',r),(d,d',s) \big) \mapsto \big( (c,d),(c',d'),(r,s) \big)
	\end{equation*}
	and on morphisms as
	\begin{equation*}
		\big( (f,f'),(g,g') \big) \mapsto \big( (f,g),(f'g') \big).
	\end{equation*}
	Functoriality and $\ast$-commutation are immediate.
	Writing $1$ for the terminal CSMC, with one object $\bullet$, it is clear that $\Chu(1,\bullet)$ has only one object and is isomorphic to the terminal $\ast$-autonomous category.
	The required 2-cells (see e.g.\cite{day_monoidal_bicats}) are straightforward to define and are all isomorphisms.
\end{IEEEproof}

\subsection{Proof of Theorem \ref{thm:pseudomonoids}}\label{app:pseudomonoids}

\begin{IEEEproof}
	A pseudomonoid in $\staraut$ is a $\ast$-autonomous category $\cat{C}$ equipped two $\ast$-functors $-\seq -:\cat{C}\times\cat{C}\morph{}\cat{C}$ and $i_\seq:1\morph{}\cat{C}$ and natural isomorphisms making these essentially associative and unital.
	Since $\ast$-functors are equivalent to degenerate linear functors between $\ast$-autonomous categories by Theorem \ref{thm:star_functor}, we see that we recover the definition of a 
	pre-BV-category given in \cite{blute_BV}.

	The normality of functors in $\nstaraut$ enforces the extra condition $i_\otimes\cong i_\seq$ making a pseudomonoid here a BV-category.
\end{IEEEproof}

\subsection{Proof of Theorem \ref{thm:preBV}}\label{app:preBV}
\begin{IEEEproof}
	By Proposition \ref{prop:chu_monoidal}, $\Chu$ is a strong monoidal 2-functor.
	Any (even lax) monoidal $2$-functor $F: \cat{C} \rightarrow \cat{D}$ preserves pseudomonoids \cite{day_monoidal_bicats}, and furthermore lifts along the forgetful functors $U$ to a $2$-functor $F:\text{PsMon}(\cat{C}) \rightarrow \text{PsMon}(\cat{D})$ such that the following diagram commutes \cite{McCrudden2000Balanced}.
	\[
		\begin{tikzcd}
			\text{PsMon}(\cat{C}) \arrow[rr, "\text{PsMon}(F)"] \arrow[d, "U"'] & & \text{PsMon}(\cat{D}) \arrow[d, "U"] \\
			\cat{C} \arrow[rr, "F"'] & & \cat{D}
		\end{tikzcd}
	\] 
	From now on we identify $\Chu$ with $\text{PsMon}(\Chu)$ whenever the context is clear.
	So, if $(\cat{C},\bot)$ is a pseudomonoid in $\mathsf{CSMC}^{\lrcorner}_\bot$ then $\Chu(\cat{C},\bot)$ is automatically a pseudomonoid in $\staraut$, that is, a pre-BV-category by Theorem \ref{thm:pseudomonoids}. 

	Now consider a pseudomonoid in $\mathsf{CSMC}^{\lrcorner}_\bot$.
	This consists of a multiplication and unit of the form,
	\begin{equation*}
		\seq:(\cat{C},\bot)\times(\cat{C},\bot) \morph{} (\cat{C},\bot), \hspace{1cm} i_\seq:(1,\bullet) \morph{} (\cat{C},\bot).
	\end{equation*}
	This makes $\cat{C}$ a $\otimes$-closed, $\otimes$-symmetric duoidal category with all pullbacks.
	Since $\seq$ must laxly preserve the chosen objects, we have morphisms $\bot\seq\bot\morph{}\bot$ and $i_\seq\morph{}\bot$ making $\bot$ a $\seq$-monoid; the associativity and unitality come from those of the pseudomonoid.
\end{IEEEproof}

\subsection{Proof of Theorem \ref{thm:BV}}\label{app:BV}
\begin{IEEEproof}
	The proof proceeds much like the previous one.
	By Proposition \ref{prop:chu_monoidal}, $\Chu$ is strongly monoidal and therefore sends pseudomonoids to pseudomonoids.
	So, if $(\cat{C},\bot)$ is a pseudomonoid in $\mathsf{CSMC}^{\lrcorner}_{\mathsf{N},\bot}$ then $\Chu(\cat{C},\bot)$ is a pseudomonoid in $\staraut_{\mathsf{N}}$, that is, a BV-category. 

	Now consider a pseudomonoid on $(\cat{C},\bot)$ in $\mathsf{CSMC}^{\lrcorner}_{\mathsf{N},\bot}$.
	This makes $\cat{C}$ a $\otimes$-closed, $\otimes$-symmetric normal duoidal category with all pullbacks.
	Now $\seq$ must preserve the chosen objects, giving isomorphisms $\bot\seq\bot\cong\bot$ and $i_\seq\cong\bot$.
	The latter, along with normality of the duoidal structure implying that $\bot\cong i_\otimes$.
\end{IEEEproof}

\subsection{Proof of Theorem \ref{thm:superunitaries}}\label{app:superunitaries}

\begin{IEEEproof}
	As outlined in the main text, morphisms of type $ \otimes_i( \cat{C}_{\vb*{a}_i},y_{\vb*{a}_i} )\rightarrow \parr_{k} (\cat{C}_{\vb*{a}_k},y_{\vb*{a}_k})$ are strong natural transformations of the form $ \otimes_i  \cat{C}_{\vb*{a}_i} \rightarrow \C{b}$ with $\mathbf{b} : = (\otimes_k b_k, \otimes_k b_k')$ such that when all but one of their inputs are filled they return an optic.

	On the other hand, it is known that all unitary supermaps decompose locally as a comb \cite{Yokojima2021consequencesof}.
	As a result, it was proven in \cite{wilson_polycategories} that there is a polycategory $\mathsf{srep}(\cat{C})$ equivalent to $\mathsf{uQS}$ whose polymorphisms are taken to be strong natural transformations $S: \otimes_i \cat{C}_{\vb*{a}_i} \rightarrow   \C{b}$ with $\mathbf{b} := (\otimes_k b_k, \otimes_k b_k' ) $ such that when evaluated on all but one input they return a comb.
	
	All that remains is to use the equivalence between unitary optics and unitary quantum combs proven in \cite{hefford_coend}.
\end{IEEEproof}

\end{document}